\documentclass[twocolumn]{aastex702}
\usepackage{float}
\usepackage{natbib}
\usepackage{csvsimple}
\usepackage{booktabs}
\usepackage{comment}
\usepackage{amsmath}
\usepackage{cleveref}
\usepackage{xcolor}
\usepackage{hyperref}

\newcommand{\rev}[1]{#1}

\renewcommand{\figsetstart}{}
\renewcommand{\figsetnum}[1]{}
\renewcommand{\figsettitle}[1]{}
\graphicspath{{./plots/}}

\usepackage{booktabs} 

\begin{document}

\title{Wolf-Rayet Stars in M33: The Impact of Stellar Crowding on Spectral Classification and Mass-Loss Diagnostics}

\author[orcid=0009-0000-0170-6302, gname=Rohit, sname=Raj]{Rohit Raj}
\affiliation{Center for Astrophysics $|$ Harvard \& Smithsonian, 60 Garden Street, Cambridge, MA 02138, USA}
\email[show]{rohit.raj@cfa.harvard.edu}

\author[0000-0001-8867-4234]{Puragra Guhathakurta}
\affiliation{Department of Astronomy \& Astrophysics, University of California Santa Cruz, 1156 High Street, Santa Cruz, CA 95064, USA}
\email{raja@ucolick.org}

\author[0000-0002-3742-8460]{Robert E. Williams}
\affiliation{Space Telescope Science Institute, 3700 San Martin Drive, Baltimore, MD 21218, USA}
\affiliation{Department of Astronomy \& Astrophysics, University of California Santa Cruz, 1156 High Street, Santa Cruz, CA 95064, USA}
\email{wms@stsci.edu}

\author[0000-0002-7502-0597]{Benjamin F.\ Williams}
\affiliation{Department of Astronomy, University of Washington, Box 351580, Seattle, WA 98195-1580, USA}
\email{benw1@uw.edu}

\author[0000-0002-1264-2006]{Julianne J.\ Dalcanton}
\affiliation{Center for Computational Astrophysics, Flatiron Institute, 162 Fifth Avenue, New York, NY 10010, USA}
\affiliation{Department of Astronomy, University of Washington, Box 351580, Seattle, WA 98195-1580, USA}
\email{jdal@uw.edu}

\author[0000-0001-7531-9815]{Meredith J.\ Durbin}
\affiliation{Department of Astronomy, University of Washington, Box 351580, Seattle, WA 98195-1580, USA}
\affiliation{Department of Astronomy, University of California, Berkeley, CA 94720, USA}
\email{meredith.durbin@berkeley.edu}

\begin{abstract}
\noindent
The metallicity dependence of Wolf-Rayet (WR) mass-loss rates is a critical input for massive star evolutionary models, chemical yield predictions, and ionizing radiation budgets of young stellar populations. The Triangulum Galaxy (M33), with 213 spectroscopically confirmed WR stars spanning a $\sim$0.5~dex radial metallicity gradient, is the premier laboratory for this calibration. However, the ground-based fiber spectra on which the current census rests blend each target with neighboring OB stars, diluting diagnostic emission-line equivalent widths and biasing spectral classifications, inferred mass-loss rates, and surface hydrogen abundances. Using $0\farcs1$-resolution \textit{HST}/PHATTER survey imaging, we quantify this contamination for 123 WR stars within the survey footprint, finding that $\sim$27\% suffer moderate-to-severe dilution from unresolved neighbors with WR continuum flux contributions $F_{\rm WR} < 0.85$. We provide crowding classifications, $F_{\rm WR}$ fractions, and six-band photometry in the F275W, F336W, F475W, F814W, F110W, and F160W bands for each identified counterpart. We also present Keck/DEIMOS slit spectra of five WR stars, which are substantially less contaminated than the discovery MMT/Hectospec fiber spectra. A direct equivalent-width comparison of these targets confirms the predicted dilution. The slit spectra yield \rev{blue-edge velocities of \ion{He}{1} absorption ($v_{\rm blue} \approx 700$--$1500$~km\,s$^{-1}$)}, comparable to WR stars at similar metallicity in the LMC. The $F_{\rm WR}$ correction factors, together with the slit-spectrum benchmarks, provide the inputs needed to place M33 on the $\dot{M}$--$Z$ relation as a calibration anchor between the LMC and the Galaxy.
\end{abstract}


\section{Introduction}\label{sec:intro}

Massive stars have brief yet influential lifetimes, and significantly impact the interstellar medium (ISM) through powerful stellar winds, radiation, and their ultimate fate as core-collapse supernovae \citep{Massey85, Smith14, Crowther07}. Wolf-Rayet (WR) stars represent a particularly consequential evolutionary phase, characterized by \rev{mass-loss rates spanning $\dot{M} \sim 10^{-6}$--$10^{-4}\,M_\odot\,\mathrm{yr}^{-1}$} and helium-rich surfaces that significantly shape galactic chemical enrichment \citep{Conti84, Crowther07, Meynet05, Hainich15, Hamann19}. \rev{At approximately Galactic and Large Magellanic Cloud (LMC) metallicities, WR stars may originate from O-type progenitors with initial masses above $\sim$20--25\,$M_\odot$ \citep{Shenar20}} and occupy the WR phase for roughly the final 10\% of their lifetimes \citep{Meynet05}. They serve as markers of recent high-mass star formation and are potential progenitors of stellar-mass black holes and long-duration gamma-ray bursts \citep{Woosley06, Hirschi05}. \rev{The defining observational characteristic of WR stars is their broad emission-line-dominated spectra, produced by dense, fast stellar winds \citep{Crowther07}.} Two primary channels produce WR-type spectra: very massive main-sequence stars ($\gtrsim 80$--$100\,M_\odot$) that retain hydrogen-rich envelopes (WNh stars), and classical WR (cWR) stars---core helium-burning, hydrogen-depleted objects with $T_\mathrm{eff} > 30{,}000$~K that constitute $\sim$90\% of the known population \citep{Grafener08, Shenar21, Shenar24}.
 
Observations of WR stars provide critical constraints on massive star evolutionary models, particularly the poorly understood interplay between mass loss, rotation, and metallicity \citep{Chiosi86, Smith14, Ou23}. In the Milky Way, non-LTE atmospheric analyses with codes such as \textsc{cmfgen} \citep{Hillier98} and \textsc{PoWR} \citep{Hamann06} have yielded relatively precise wind parameters for individual WR stars \citep{Hamann19, Sander19}, but the narrow metallicity range within the Galaxy limits tests of the $\dot{M}$--$Z$ dependence. Surveys in the Magellanic Clouds have extended this baseline to sub-solar metallicities \citep{Hainich14, Shenar19}, and population synthesis studies incorporating massive binaries \citep{Eldridge09} have demonstrated the importance of binary channels. However, fully constraining the metallicity dependent changes requires measuring the characteristics of WR populations spanning a broad range of chemical environments, both within galaxies with known abundance gradients and across samples of galaxies at different mean metallicities. 

The nearby Triangulum galaxy (M33) provides such a laboratory. At a distance of $\sim$840~kpc \citep{Freedman91, Lee22}, it is close enough for individual massive stars to be resolved as well as a single survey can capture a statistically significant WR population spanning a range of galactic environments. Its relatively face-on disk orientation ($i \approx 56\degr$; \citealt{Regan94}) minimizes differential extinction along the line of sight, while its high star-formation rate and large luminosity ensure a rich massive star population \citep{Neugent11, Cullinane23}. Most critically, M33 hosts a well-characterized radial metallicity gradient spanning $\sim$0.5~dex from the nucleus to the outer disk \citep{Rosolowsky08, Bresolin11}, enabling a direct examination of how chemical environment shapes WR star properties such as wind velocities and mass-loss rates \citep{Cervi99, JDSimon08, Ramachandran19}. The spectroscopic census by \citet{Massey85}, \citet{Neugent11, Neugent14}, and \citet{Massey21}---conducted primarily with the fiber-fed Hectospec spectrograph on the 6.5~m Multiple Mirror Telescope (MMT) \citep{Fabricant05}---has identified 213 WR stars in M33, establishing the subtype distribution and spatial demographics across the disk.

In the course of the Triangulum Extended (TREX) spectroscopic survey of M33 \citep{Gilbert22, Quirk22, Cullinane23}, we used the Keck/DEIMOS spectrograph to obtain spectra of $\approx$7000 resolved stars across the galaxy's disk and halo. TREX targets in M33's inner disk were preselected from the \textit{HST} Panchromatic Hubble Andromeda Treasury: Triangulum Extended Region (PHATTER) imaging survey \citep{Williams21} to be relatively isolated point sources. During our systematic inspection of these spectra, we found five previously known WR stars as serendipitous detections. Because the targets were selected for isolation and observed in sub-arcsecond seeing ($\lesssim 0\farcs8$)
\rev{through $0\farcs8$--$1\farcs0$-wide slits},
these slit spectra are substantially less contaminated by
neighboring sources than the original Hectospec fiber spectra
\rev{($1\farcs5$ diameter; observed under typical
$\sim$\,$1\arcsec$ seeing at the MMT; \citealt{Massey16})}
from which the WR classifications were derived.

This serendipitous discovery, and the visibly cleaner spectral features it revealed, motivated us to carry out a systematic analysis of the broader WR population within the PHATTER footprint. At M33's distance, a typical $1\arcsec$ seeing disk corresponds to $\sim$4~pc --- comparable to the typical size of compact OB associations --- such that the $1\farcs5$ Hectospec fibers \rev{sometimes} blend the WR target with \rev{nearby} massive neighbors. These unresolved OB companions dilute diagnostic emission lines (e.g., \ion{He}{2}~$\lambda4686$, \ion{C}{4}~$\lambda5808$) by contributing continuum flux, reducing measured equivalent widths (EWs) and potentially biasing spectral classifications and mass-loss rate determinations \citep{Zinnecker07, Bibby10, Crowther07}. An additional source of contamination is narrow nebular H$\alpha$ emission from the diffuse ISM, which can be captured by fiber apertures without the spatial information needed to separate it from the broad stellar wind profile, artificially inflating the apparent H$\alpha$ luminosity and hence the inferred $\dot{M}$ \citep{Crowther07}. As we demonstrate in \S\,\ref{sec:Halpha_decontam}, long-slit spectrographs such as Keck/DEIMOS \rev{partially} mitigate this by enabling spatial subtraction of the extended nebular component.
 
Using $0\farcs1$-resolution \textit{HST} imaging from the PHATTER survey \citep{Williams21}, we quantify the flux contribution from neighboring sources for 123 of the 213 M33 WR stars within the survey footprint, classify their crowding environments, and provide robust six-band photometry. We also present multi-epoch Keck/DEIMOS spectra for three of the five serendipitously observed WR stars, spanning a 4--5~year baseline. Together, the crowding-corrected photometric catalog and the new slit spectra provide a vital reference for future observational programs, population synthesis calibrations, and the development of machine learning classifiers.

This paper is organized as follows. In \S\,\ref{sec:data}, we describe the observational datasets: the M33 WR catalog, the PHATTER imaging, and the Keck/DEIMOS spectroscopy. In \S\,\ref{sec:complex_wrs}, we present the crowding analysis methodology and color-magnitude diagram results. In \S\,\ref{sec:Keck_WRs}, we analyze the serendipitous spectra, including nebular H$\alpha$ decontamination and temporal variability. In \S\,\ref{sec:discussion}, we discuss implications for spectral classification, mass-loss rates, and population synthesis. In \S\,\ref{sec:summary}, we summarize the main results.

\begin{figure*}[htb!]
\centering
\includegraphics[width=\textwidth]{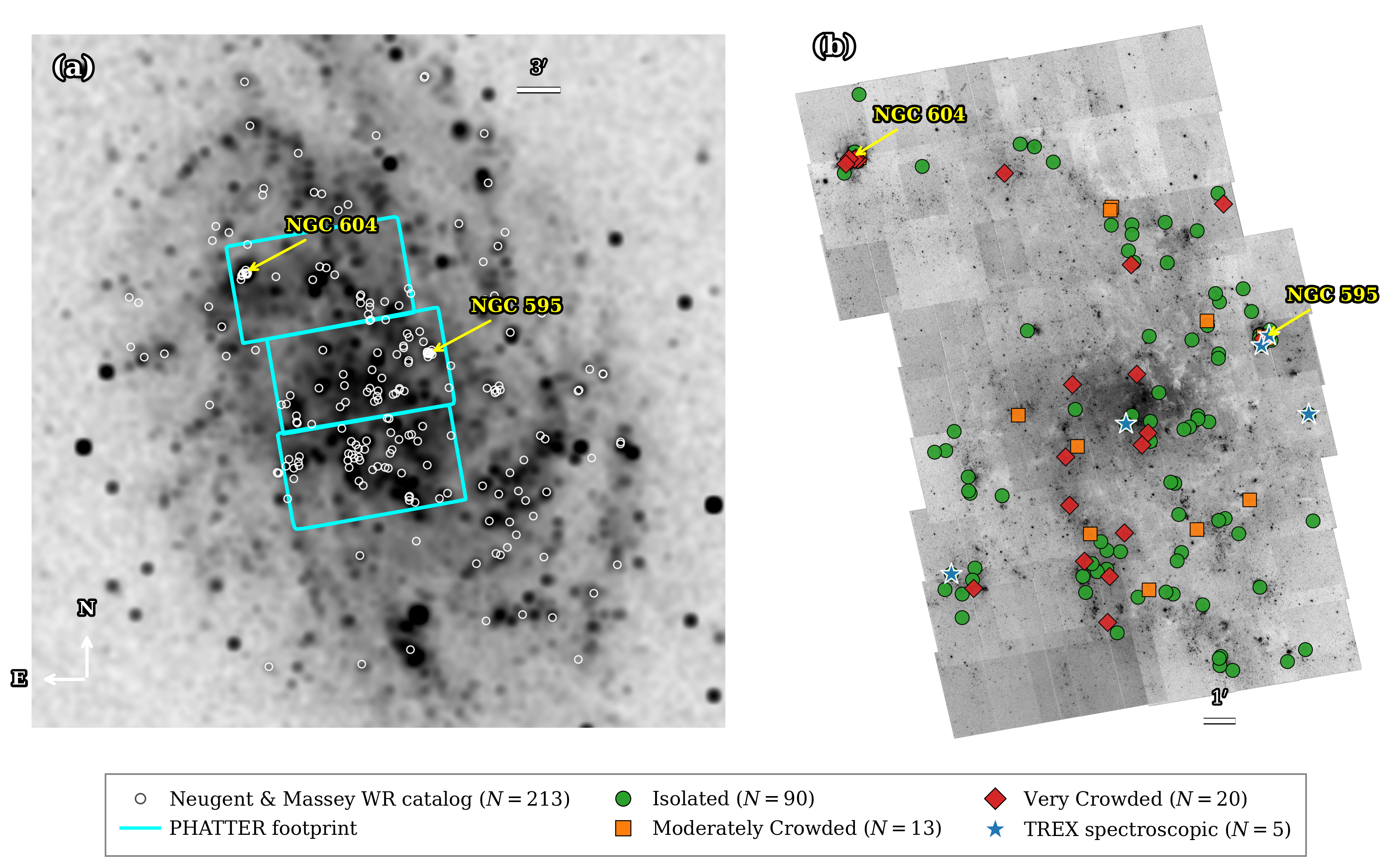}
\caption{Spatial distribution of the Wolf-Rayet population 
in M33.
\textbf{(a)}~Full-galaxy view in a GALEX near-ultraviolet (NUV) image. Prominent star formation is visible along the spiral arms. Open white circles indicate the positions of all 213 spectroscopically confirmed WR stars from \citet{Massey85}, \citet{Neugent11,Neugent14}, and \citet {Massey21}. The cyan outline indicates the 
\textit{HST}/PHATTER survey footprint \citep{Williams21}, 
which covers the inner high-surface-brightness disk and 
encloses 123 of the 213 known WR stars. Thick cyan lines denote the boundaries of the three major survey bricks. The giant \ion{H}{2} regions NGC~604 and NGC~595 are labeled.
\textbf{(b)}~Zoomed in view of the PHATTER footprint using the \textit{HST}/WFC3 F336W mosaic as background, with WR 
stars color-coded by crowding classification: isolated 
(green circles; $N = 90$), moderately crowded (orange 
squares; $N = 13$), and very crowded (red diamonds; 
$N = 20$), as defined in \S\,\ref{sec:crowding_metric}. 
Blue stars mark the five WR targets with Keck/DEIMOS 
spectra from the TREX survey (\S\,\ref{sec:Keck_WRs}). 
The very crowded WR stars cluster preferentially along 
the spiral arm and in the vicinity of NGC~604 
and NGC~595, confirming that the photometric contamination 
quantified in this work is a systematic effect driven by 
the dense stellar environments associated with active 
star formation.}
\label{fig:M33_WRs_Location}
\end{figure*}

\section{Data}\label{sec:data}
\rev{This study uses the published spectroscopic catalog of M33
WR stars (\S\,\ref{sec:M33_WRs}) to define the parent sample,
together with two observational datasets: \textit{HST}/PHATTER
imaging (\S\,\ref{sec:hst_phot}) and Keck/DEIMOS slit spectroscopy
from the TREX survey (\S\,\ref{sec:keck_spec}).}

\subsection{\rev{The M33 Wolf-Rayet Parent Sample}}\label{sec:M33_WRs}
The parent sample for our study is the comprehensive catalog of WR stars in M33, which is the result of several foundational spectroscopic surveys carried out using the 6.5-m MMT/Hectospec multi object fiber spectrograph \citep{Fabricant05}. The primary census by \citet{Neugent11} listed 206 WR stars and an additional six candidates. Subsequent work by \citet{Neugent14} confirmed five of these candidates as bona fide WR stars. Including another confirmation by \citet{Massey16} and one new discovery listed in \citet{Massey21}, the total known population of spectroscopically identified WR stars in M33 currently stands at 213.
We note that this census includes objects with complex classifications. For instance, the star J013418.37+303837.0 (Var 2) is a known Luminous Blue Variable (LBV), but it exhibits an Ofpe/WN9 spectrum and is often included in WR population counts \citep{Massey21}.
The issue of stellar crowding for M33 WR stars has been previously considered, particularly in the dense cores of the NGC 604 and NGC 595 young clusters \citep{Massey98, Neugent11}, and surveyed using HST by \citet{Drissen93}.

\subsection{Hubble Space Telescope Photometry from the PHATTER Survey}\label{sec:hst_phot}
The M33 photometry is drawn from the PHATTER survey \citep{Williams21},
a 108-orbit \textit{HST} program that imaged $\sim$22 million resolved
stars across the inner disk in six filters: F275W and F336W in the near ultraviolet (the latter is similar to ground based $U$) using Wide Field Camera 3 ultraviolet/visible light (WFC3/UVIS), F475W and F814W in the optical (similar to ground based $B$ and $I$, respectively) using Advanced Camera for Surveys wide field (ACS/WFC), and F110W and F160W in the near infrared (similar to ground based $J$ and $H$, respectively) using WFC3 near-infrared (WFC3/IR). The survey footprint comprises three contiguous bricks covering an approximately $12' \times 18'$ region; full details of the tiling strategy, photometric pipeline, and forced PSF-fitting
catalogs are given by \citet{Williams21}. The catalogs are publicly
available via MAST.\footnote{\url{https://archive.stsci.edu/hlsp/phatter}}

Of the 213 known M33 WR stars, 123 fall within the PHATTER footprint.
These 123 targets form the basis of our crowding analysis
(\S\,\ref{sec:complex_wrs}). \rev{The $0\farcs1$-resolution PHATTER
imaging enables the first systematic, star-by-star crowding
characterization of this population.}

\subsection{Keck DEIMOS Spectroscopy}\label{sec:keck_spec}
Spectroscopic data were obtained using the DEep Imaging Multi-Object Spectrograph (DEIMOS) mounted on the Keck II 10-meter telescope at the summit of Mauna Kea, Hawaii. DEIMOS, developed by \citet{Faber03}, provides medium spectral resolution and was an integral part of the TREX, which targeted stars across M33's disk and halo to explore its stellar kinematics \citep{Gilbert22, Quirk22, Cullinane23}. A total of $\approx7000$ stars were observed during the TREX survey, making it currently the largest stellar spectroscopic survey of M33, spanning the galaxy's entire disk out to a deprojected radius of $\sim$11~kpc \citep{Gilbert22, Quirk22, Cullinane23}. Of the 123 WR stars within the PHATTER footprint, five WR stars were captured by TREX as serendipitous detections (\S\,\ref{sec:Keck_WRs}).

Spectroscopic targets were selected from the M33 {\it HST\/} PHATTER survey. The DEIMOS observations were obtained using two gratings: (1)~600ZD (600~line~mm$^{-1}$, $R \approx 2000$) for spectra targeting young blue stars, and (2)~1200G (1200 line mm$^{-1}$, $R \approx 6000$) for older, red giant branch (RGB) stars. The spectral coverage is $\lambda\lambda\sim4600$--9800\,\AA\ and $\lambda\lambda\sim6300$--9800\,\AA\ for the 600ZD and 1200G gratings, respectively.

Data were reduced with the \texttt{spec2d} and \texttt{spec1d} pipelines \citep{cooper2012spec2d, Newman13}, which perform flat-fielding, sky subtraction, wavelength calibration, and slit-centering corrections. Heliocentric radial velocities were measured by cross-correlating the observed spectra with stellar templates \citep{Tonry79}.

\begin{figure*}[htb!]
\centering
\includegraphics[width=\textwidth]{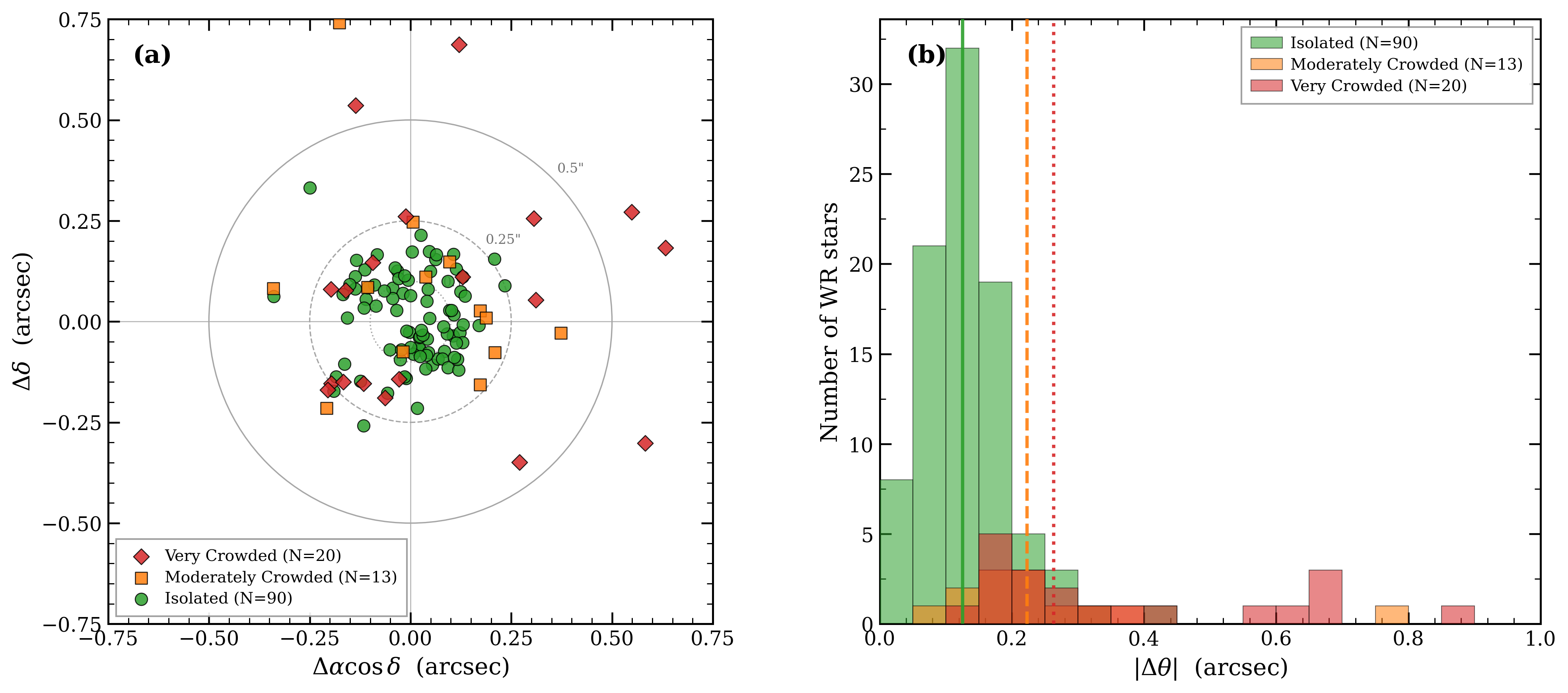}
\caption{Astrometric residuals between ground-based catalog positions and \textit{HST}/PHATTER counterparts for 123 M33 WR stars 
within the PHATTER footprint, color-coded by crowding classification.  
A bulk systematic offset between the ground-based astrometric frame 
and the \textit{HST}/PHATTER reference frame has been removed prior 
to this comparison (\S\,\ref{sec:crossmatch}).
\textbf{(a)}~Vector-point diagram showing $\Delta\alpha\cos\delta$ 
vs.\ $\Delta\delta$, the residual offset components in right ascension 
and declination.  Reference circles at $0\farcs10$, $0\farcs25$, and 
$0\farcs50$ are shown for scale.  Green circles, orange squares, and 
red diamonds denote the isolated ($N = 90$), moderately crowded 
($N = 13$), and very crowded ($N = 20$) samples, respectively.  
Isolated sources cluster tightly around the origin, while very crowded 
sources extend to larger, more isotropically distributed offsets.
\textbf{(b)}~Histogram of the total angular offset $|\Delta\theta|$ 
for each crowding class; vertical lines mark the class medians 
(isolated: $0\farcs13$; moderately crowded: $0\farcs22$; very crowded: 
$0\farcs26$, maximum $0\farcs88$).  The progressive broadening of 
the offset distribution from isolated to very crowded is a direct, 
model-independent signature of ground-based centroid displacement 
by unresolved blended neighbors, consistent with the 
crowding classifications derived from the two-parameter photometric 
metric (Figure~\ref{fig:crowding}).}
\label{fig:astrometric_offsets}
\end{figure*}

\section{Complex Environment of Wolf-Rayet Stars}\label{sec:complex_wrs}
The primary goal of our analysis is to robustly identify the \textit{HST} counterparts to the ground-based M33 WR catalog and to quantify the degree of photometric contamination affecting each source. As discussed in \S\,\ref{sec:intro}, ground-based seeing blends WR targets with neighboring sources; our methodology quantifies this  contamination using the PHATTER catalog.

\subsection{Cross-Matching and Counterpart Identification}\label{sec:crossmatch}
For each of the 123 targets, we queried the PHATTER point-source
catalog \citep{Williams21} for all resolved sources within $1\arcsec$
of the published ground-based WR coordinates, a radius comparable to
the typical seeing disk of the foundational surveys
\citep{Neugent11}. We identified the WR counterpart as the brightest
F275W source within this aperture: classical WR stars ($T_{\mathrm{eff}}
\gtrsim 30{,}000$~K; \citealt{Crowther07, Shenar24}) emit predominantly
in the ultraviolet, providing maximum contrast against cooler OB
neighbors.

Using this criterion, we securely identified a unique, bright F275W counterpart for 113 of the 123 cases. The median F275W magnitude of \rev{the identified counterparts} is 18.6 (interquartile range 17.0--19.2), well above the PHATTER 50\% completeness limit \rev{($m_{\rm F275W} \approx 24.1$--$24.4$ across all stellar density bins \citep{Williams21})}. At this brightness the surface density of
PHATTER sources ($\rho \approx 1.4 \times 10^{-2}$~arcsec$^{-2}$)
yields a chance-superposition probability within $1\arcsec$ of only
$\sim$4\%. The remaining 10 targets possess only faint counterparts ($m_{\mathrm{F275W}} \gtrsim 23$). Because the chance alignment probability at these faint magnitudes rises above $\sim$40\%, we flag these $\sim$10 targets as ambiguous and discuss them separately in \S\,\ref{sec:amb_cases}.

\subsection{Synthetic Fiber Aperture Model}\label{sec:synth_fiber}
To simulate the ground-based spectroscopic observations from which the parent WR catalog was constructed, we developed a synthetic fiber aperture analysis. For each target, we centered a circular aperture of radius $r_{\mathrm{fib}} = 0\farcs75$ on the published WR coordinates, matching the $1\farcs5$-diameter fibers of the Hectospec multi-object spectrograph on the 6.5~m MMT \citep{Fabricant05}. We queried all PHATTER sources within $2\farcs87$ ($5\sigma$ of the adopted seeing model; see below) of the catalog coordinates and modeled the ground-based flux distribution of each source as a circular two-dimensional Gaussian, \begin{equation}\label{eq:gaussian_psf} I(r) = I_0 \, \exp\!\left( -\frac{r^2}{2\sigma^2} \right), 
\end{equation} 
where $r$ is the angular distance from the source centroid and $\sigma = \mathrm{FWHM}/(2\sqrt{2\ln 2})$. We adopted $\mathrm{FWHM} = 1\farcs0$, representative of median ground-based seeing at the observatories used for the foundational surveys \citep{Neugent11, Massey21}. This Gaussian approximation is standard practice for crowding simulations in dense stellar fields \citep{Mighell05}. For each source $i$ at angular offset $d_i$ from the fiber center, the fraction of its total flux captured by the fiber aperture is 
\begin{multline}\label{eq:fiber_fraction}
    f_i = \frac{1}{2\pi\sigma^2}
    \iint_{x^2+y^2 \,\leq\, r_{\mathrm{fib}}^2} \\
    \exp\!\left(-\frac{(x-x_i)^2 + (y-y_i)^2}
    {2\sigma^2}\right) dx\,dy,
\end{multline}
where $(x_i, y_i)$ is the offset of source $i$ from the fiber center in arcseconds. This integral is evaluated numerically on a fine grid ($0\farcs005$~pixel$^{-1}$); for a centered source ($d_i = 0$) it reduces analytically to $1 - \exp(-r_{\mathrm{fib}}^2/2\sigma^2)$, which we use as a sanity check. The fractional flux contribution of source $i$ to the total fiber-captured flux is then 
\begin{equation}\label{eq:frac_contrib} F_i = \frac{f_i \cdot 10^{-0.4\,m_i}} {\displaystyle\sum_{j} f_j \cdot 10^{-0.4\,m_j}}, 
\end{equation} 
where the sum runs over all PHATTER sources within the search radius and $m_i$ is the apparent magnitude of source $i$ in the relevant filter. For the identified WR counterpart, $F_{\rm WR} \equiv F_i$ provides a physically motivated, model-dependent measure of contamination that complements the model-independent crowding metric defined in \S\,\ref{sec:crowding_metric}. We evaluated Equation~\ref{eq:frac_contrib} independently in the F275W, F336W, F475W, and F814W filters. $F_{\rm WR}$ decreases monotonically from UV to optical wavelengths (median values: 0.90, 0.88, 0.69, and 0.46, respectively), reflecting the increasing relative contribution of cooler neighbors at longer wavelengths. The difference $\Delta F_{\rm WR} \equiv F_{\rm WR}^{275} - F_{\rm WR}^{814}$ has a median of 0.28 with an interquartile range of 0.11--0.47 across our 123-star sample; 77\% of targets show $\Delta F_{\rm WR} > 0.10$, confirming that the wavelength dependence is not negligible for the majority of the sample. This has two practical consequences. For crowding \textit{classification} and counterpart identification, we adopt the F275W band, where the contrast between hot WR continua and cooler neighbors is greatest (\S\,\ref{sec:crossmatch}). For quantitative \textit{EW corrections} of optical spectral diagnostics, however, the appropriate $F_{\rm WR}$ is that evaluated at the wavelength of the emission line: the tabulated F475W-band value for each star applies to lines at $\sim$4500--5500\,\AA\ (e.g., \ion{He}{2}\,$\lambda$4686, \ion{He}{2}\,$\lambda$5411), and the F814W-band value applies to lines at $\gtrsim$7000\,\AA. Both filter-band $F_{\rm WR}$ values are provided in the machine-readable catalog.

The physical interpretation of $F_{\rm WR}$ as a contamination measure is supported by the residual astrometry. The angular offset $d$ between the identified PHATTER counterpart and the published catalog coordinates serves as a model-independent crowding diagnostic (\S\,\ref{sec:crowding_metric}): for an isolated WR star the ground-based flux-weighted centroid coincides with the true stellar position ($d \approx 0$), whereas blending with a nearby comparably-bright neighbor pulls the centroid toward the secondary, increasing $d$. Figure~\ref{fig:astrometric_offsets} quantifies this effect. After applying a bulk astrometric correction to place the ground-based catalog on the \textit{HST}/PHATTER reference frame (\S\,\ref{sec:crossmatch}), the residual positional offsets show a clear dependence on crowding class: isolated WR stars have a median residual of $0\farcs13$ ($\sigma \approx 0\farcs10$), consistent with the intrinsic astrometric precision of the ground-based surveys \citep{Neugent11}, while moderately and very crowded sources show systematically larger residuals---medians of $0\farcs22$ and $0\farcs26$, respectively, with scatter increasing to $\sigma \approx 0\farcs19$--$0\farcs33$---consistent with flux-weighted centroid displacement toward unresolved bright neighbors. The maximum offset among very crowded stars reaches $0\farcs88$.

\begin{figure}[htb!]
\centering
\includegraphics[width=\columnwidth]{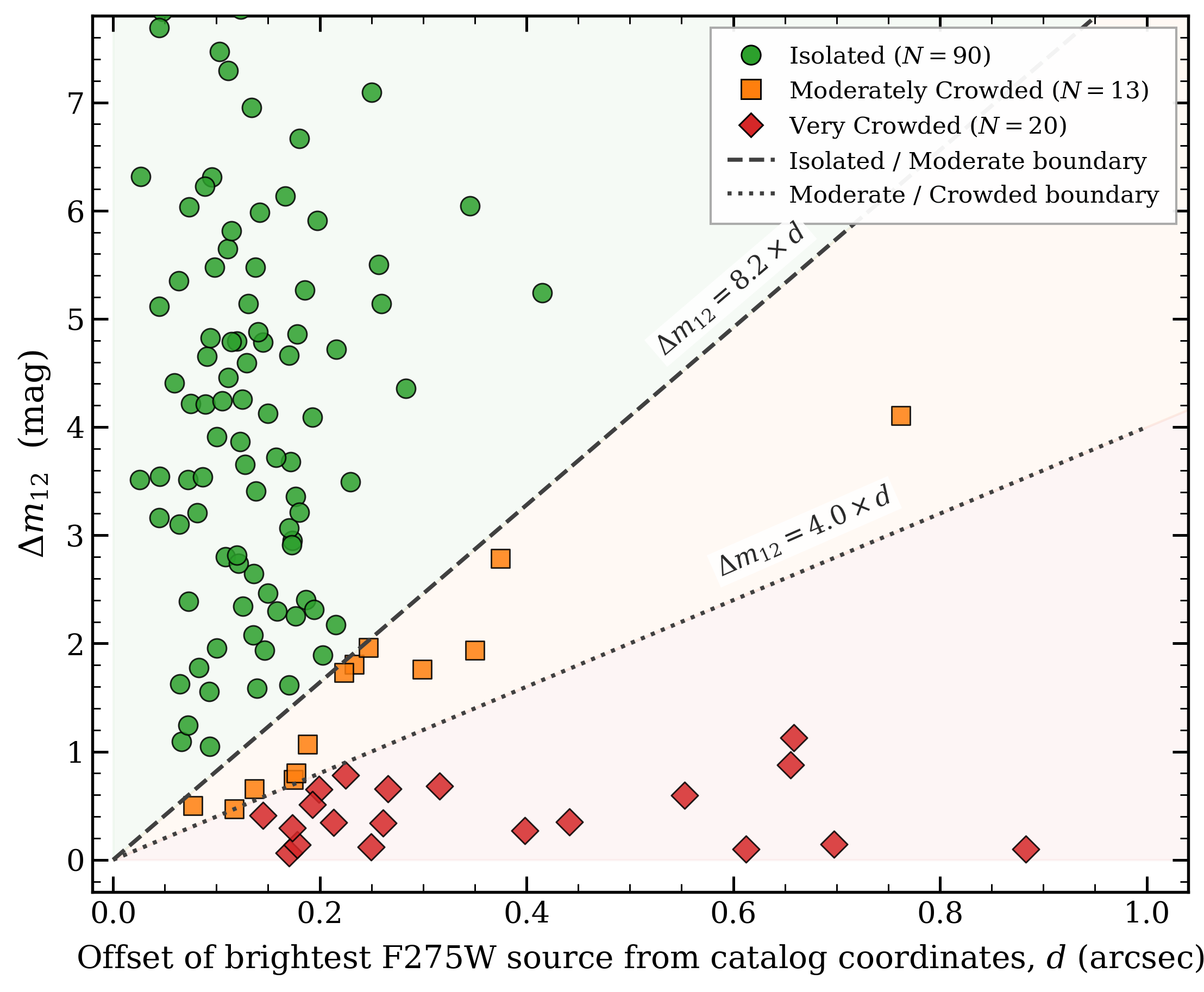}
\caption{Two-parameter crowding diagnostic for 123 M33 WR stars within the PHATTER footprint, evaluated in the F275W band. The horizontal axis shows the angular offset $d$ of the brightest F275W PHATTER source from the published ground-based WR coordinates; the vertical axis shows the magnitude difference $\Delta m_{12}$ between the brightest and second-brightest F275W sources within $1\arcsec$. The dashed and dotted lines show the empirical classification boundaries $\Delta m_{12} = 8.2\times d$ (isolated/moderately crowded) and $\Delta m_{12} = 4.0\times d$ (moderately crowded/very crowded), respectively (\S\,\ref{sec:crowding_metric}). Green circles, orange squares, and red diamonds mark the isolated, moderately crowded, and very crowded samples. Over one quarter of the sample (moderately + very crowded) has ground-based photometry and spectral classifications potentially compromised by unresolved continuum dilution.}
\label{fig:crowding}
\end{figure}

\begin{deluxetable*}{lccccccccr}
\tablecaption{Catalog of M33 Wolf-Rayet Stars with Crowding Analysis\label{tab:wr_catalog}}
\tablewidth{0pt}
\tablehead{
\colhead{IAU Name} & \colhead{R.A.\tablenotemark{a}} & \colhead{Decl.\tablenotemark{a}} & \colhead{Cross-ID} & \colhead{Spectral} & \colhead{Crowding} & \colhead{$d$\tablenotemark{b}} & \colhead{$\Delta m_{12}$\tablenotemark{c}} & \colhead{$F_{\rm WR}$\tablenotemark{d}} & \colhead{Coord.} \\
\colhead{} & \colhead{(deg)} & \colhead{(deg)} & \colhead{} & \colhead{Type} & \colhead{Category} & \colhead{(arcsec)} & \colhead{(mag)} & \colhead{} & \colhead{Source}
}
\startdata
M33 J013326.57+303550.1 & 23.360724 & 30.597252 & \nodata & WN6 & Isolated & 0.128 & 3.66 & 0.82 & PHATTER \\
M33 J013333.26+304129.6 & 23.388598 & 30.691548 & AM 5 & WC6 & Moderately Crowded & 0.175 & 0.74 & 0.41 & PHATTER \\
M33 J013333.79+304129.7 & 23.390800 & 30.691570 & N595-WR9 & WN7 & Very Crowded & 0.249 & 0.12 & 0.32 & PHATTER \\
M33 J013340.18+303551.6 & 23.417414 & 30.597660 & MC 40 & WC4-5 & Isolated & 0.129 & 4.59 & 0.84 & PHATTER \\
M33 J013343.30+303533.9 & 23.430433 & 30.592747 & UIT 177 & WN4.5+O6-9 & Moderately Crowded & 0.117 & 0.47 & 0.46 & PHATTER \\
M33 J013348.81+303949.6 & 23.453365 & 30.663783 & MJ X4 & WN & Isolated & 0.170 & 1.62 & 0.47 & PHATTER \\
M33 J013353.78+303528.5 & 23.474080 & 30.591256 & MJ G34 & WNE & Very Crowded & 0.178 & 0.14 & 0.29 & PHATTER \\
M33 J013358.67+303525.7 & 23.494460 & 30.590461 & OB 2-4 & B1 Ia+WNE & Moderately Crowded & 0.762 & 4.10 & 0.73 & PHATTER \\
M33 J013404.03+304658.2 & 23.516787 & 30.782825 & \nodata & \nodata & Isolated & 0.106 & 4.24 & 0.94 & PHATTER \\
M33 J013411.12+304636.9 & 23.546322 & 30.776907 & \nodata & WC4 & Very Crowded & 0.553 & 0.59 & 0.34 & PHATTER \\
\enddata
\tablenotetext{}{This table is published in its entirety (213 entries) in machine-readable format. A portion is shown here for guidance regarding its form and content.}
\tablenotetext{a}{J2000 coordinates. Where the star falls within the PHATTER footprint, coordinates are those of the identified brightest F275W counterpart (Coord.\ Source = ``PHATTER"); otherwise they are from \citet{Neugent11}.}
\tablenotetext{b}{Angular offset of the brightest F275W PHATTER source from the published ground-based WR coordinates (\S\,\ref{sec:crowding_metric}).}
\tablenotetext{c}{F275W magnitude difference between the brightest and second-brightest PHATTER sources within $1\arcsec$ of the catalog position. A large value indicates the WR counterpart dominates the local flux.}
\tablenotetext{d}{Fractional WR continuum contribution $F_{\rm WR}$ within a synthetic $0\farcs75$-radius fiber aperture centered on the catalog coordinates (\S\,\ref{sec:synth_fiber}), evaluated in the F275W band for crowding classification (\S\,\ref{sec:crowding_metric}). Values near 1.0 indicate minimal contamination; values $\ll 1$ indicate severe dilution from neighboring sources. For EW corrections of optical emission lines, F475W- and F814W-band $F_{\rm WR}$ values---typically 10--40\% lower than the F275W values---are provided in the machine-readable catalog.}
\end{deluxetable*}

\subsection{The Crowding Metric}
\label{sec:crowding_metric}
While the fractional flux contribution $F_{\rm WR}$ 
(Equation~\ref{eq:frac_contrib}) provides a physically 
motivated measure of contamination, it requires adopting 
a specific seeing model and fiber geometry. As a 
complementary, model-independent characterization of 
source confusion, we developed a two-parameter crowding 
diagnostic based on quantities measured directly from 
the PHATTER catalog.

For each WR target, we define:
\begin{itemize}
    \item $d$\,: the angular offset between the 
    published ground-based WR coordinates and the 
    identified PHATTER counterpart 
    (\S\,\ref{sec:crossmatch});
    \item $\Delta m_{12}$\,: the F275W magnitude 
    difference between the brightest and 
    second-brightest PHATTER sources within $1\arcsec$ 
    of the catalog coordinates.
\end{itemize}

\rev{These parameters provide complementary indicators
of local source confusion. A larger positional offset
$d$ indicates a greater discrepancy between the
published ground-based position and the identified
PHATTER counterpart, while a smaller $\Delta m_{12}$
indicates the presence of a comparably bright nearby
source. We therefore use their joint distribution to
characterize the degree of local crowding.}

\rev{The $(d,\Delta m_{12})$ metric is intended as an
empirical, model-independent description of source
confusion rather than as a quantitative estimate of the
fiber contamination. Larger positional offsets and
smaller magnitude contrasts both indicate greater
potential confusion, motivating a two-dimensional
partition of the observed distribution. We therefore
adopt simple linear boundaries in the
$(d,\Delta m_{12})$ plane to define qualitative
crowding categories. Quantitative contamination
estimates are instead based on the independently
calculated per-star $F_{\rm WR}$ values from the full
synthetic-fiber model (\S\,\ref{sec:synth_fiber}).}

\rev{The $(d,\Delta m_{12})$ distribution
(Figure~\ref{fig:crowding}) shows a concentration of
sources at small $d$ and large $\Delta m_{12}$ and a
tail toward larger offsets and smaller magnitude
contrasts. For a compact and reproducible description
of this continuous distribution, we define three
crowding categories using the following empirical
boundaries:}

\begin{align}
    \text{\textbf{Isolated:}} &\quad 
        \Delta m_{12} \geq 8.2 \times d, \label{eq:clean} \\[4pt]
    \text{\textbf{Moderate:}} &\quad 
        4.0 \times d \leq \Delta m_{12} < 8.2 \times d, \label{eq:moderate} \\[4pt]
    \text{\textbf{Very Crowded:}} &\quad 
        \Delta m_{12} < 4.0 \times d. \label{eq:crowded}
\end{align}

\rev{These boundaries are descriptive rather than
contours of constant $F_{\rm WR}$. The independently
computed $F_{\rm WR}$ values nevertheless show the
expected broad ordering, with isolated sources generally
having larger WR flux fractions and crowded sources
smaller values, while exhibiting substantial scatter
(Table~\ref{tab:wr_catalog}). This scatter is expected
because $F_{\rm WR}$ includes all PHATTER sources
contributing to the synthetic fiber, whereas the
$(d,\Delta m_{12})$ metric uses only the local positional
offset and the two brightest sources within $1\arcsec$.
As a sensitivity check, varying both linear slopes by
$\pm15\%$ changes the classification of only four to
five of the 123 targets, indicating that the descriptive
sample divisions are insensitive to modest changes in
the adopted boundaries.}

The physical interpretation of each category is as 
follows:
\begin{enumerate}
    \item \textbf{Isolated} ($N = 90$; 73\%): The WR 
    counterpart strongly dominates its immediate 
    neighborhood: no comparably bright source lies 
    within $1\arcsec$. For the majority of this 
    subsample, $F_{\rm WR} \gtrsim 0.85$ in the 
    synthetic fiber model (\S\,\ref{sec:synth_fiber}), 
    and spectral classifications are considered 
    reliable. However, a small number of isolated 
    targets in dense spiral-arm environments have 
    lower $F_{\rm WR}$ values (down to $\sim$0.5) due 
    to the cumulative contribution of many faint 
    sources at $1\arcsec$--$3\arcsec$ offsets that 
    fall outside the $1\arcsec$ metric search radius 
    but within the fiber aperture. For these objects, 
    the counterpart identification is confident but 
    the fiber-level EW dilution may still be 
    significant.

    \item \textbf{Moderately Crowded} ($N = 13$; 11\%): 
    A secondary source within $1\arcsec$ is bright 
    enough to have measurably displaced the 
    ground-based centroid. $F_{\rm WR}$ typically 
    falls in the range $\sim$0.40--0.85. The expected 
    EW dilution of $\sim$20--50\% is sufficient to 
    shift absolute-EW classifications by one subtype 
    in some cases \citep{Smith96, Crowther07}. The WR 
    counterpart identification remains confident.

    \item \textbf{Very Crowded} ($N = 20$; 16\%): 
    Multiple bright sources within $1\arcsec$ 
    contribute comparably to the fiber flux 
    ($F_{\rm WR} \lesssim 0.50$ in most cases). 
    Observed EWs may be diluted by factors of 
    $\gtrsim 2$; spatially resolved spectroscopy is 
    required for definitive characterization.
\end{enumerate}

\noindent \rev{The $(d,\Delta m_{12})$ classification and
$F_{\rm WR}$ characterize crowding on complementary
spatial scales: the former uses the counterpart
positional offset and the magnitude contrast of the two
brightest sources within $1\arcsec$, while $F_{\rm WR}$
includes all PHATTER sources contributing to the
synthetic fiber. Both quantities are reported in
Table~\ref{tab:wr_catalog}.}
Crowded targets are preferentially located in the spiral arms and the cores of the giant \ion{H}{2} regions NGC~604 and NGC~595 (Figure~\ref{fig:M33_WRs_Location}), where the surface density of massive stars is highest, as expected.
We note that the crowding metric evaluated in F475W yields broadly consistent classifications (Isolated: 87; Moderate: 15; Crowded: 21). The small differences arise because cooler companions contribute relatively more flux at longer wavelengths. We adopt the F275W classification as our preferred scheme throughout, consistent with the counterpart identification in \S\,\ref{sec:crossmatch}.

\begin{figure*}[htb!]
\centering
\includegraphics[width=\textwidth]{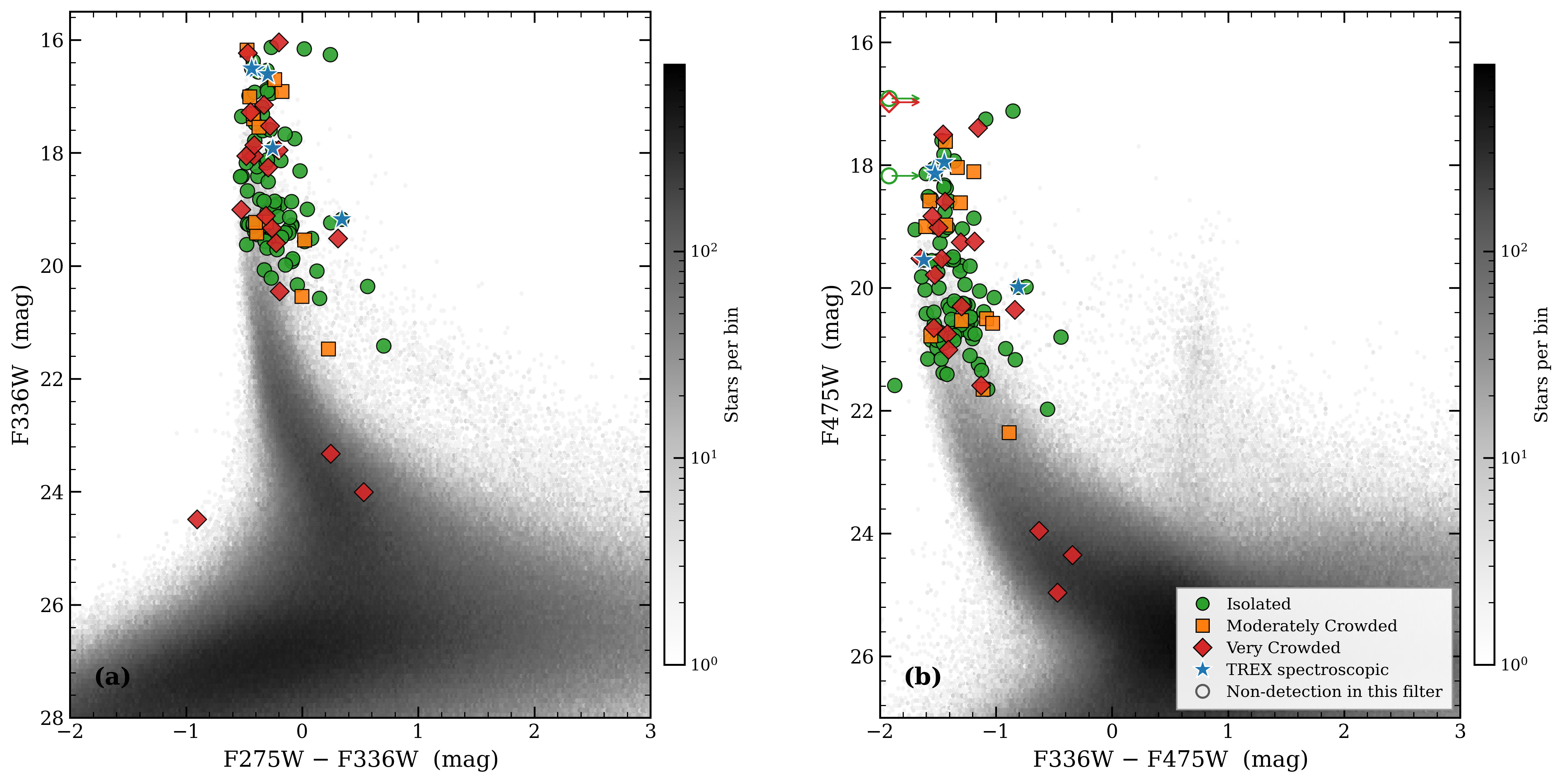}
\caption{Global color-magnitude diagrams (CMDs) of M33 from the
PHATTER survey with the WR sample overplotted, color-coded
by crowding classification.
\textbf{(a)}~NUV CMD: F336W vs.\ F275W$-$F336W.
\textbf{(b)}~Optical CMD: F475W vs.\ F336W$-$F475W.
The greyscale Hess diagram shows the density of the full
PHATTER source catalog ($\sim$22 million stars).  Green
circles, orange squares, and red diamonds denote the
isolated ($N = 90$), moderately crowded ($N = 13$), and
very crowded ($N = 20$) WR samples, respectively
(\S\,\ref{sec:crowding_metric}).  Blue stars mark the five
WR targets with Keck/DEIMOS spectra from the TREX survey
(\S\,\ref{sec:Keck_WRs}).  Three WR stars lacking valid
F336W photometry (two isolated, one very crowded) are shown
as open symbols with rightward arrows at the left boundary
of panel~(b), indicating unconstrained F275W$-$F336W
colors; they are excluded from panel~(a) where F336W enters
both axes. \rev{The three crowding categories overlap substantially in CMD space,
although a small number of very crowded targets extend to fainter
magnitudes and/or redder colors.} The remaining filter combinations
(F475W$-$F814W, F814W$-$F110W, F110W$-$F160W) are
presented in the online supplementary material.}
\label{fig:global_cmds}
\end{figure*}

\figsetstart
\figsetnum{5}
\figsettitle{Local color--magnitude diagrams for 123 M33 Wolf-Rayet stars with HST/PHATTER photometry and crowding classifications.}

\figsetend

\begin{figure*}[htb!]
\centering
\includegraphics[width=\textwidth]{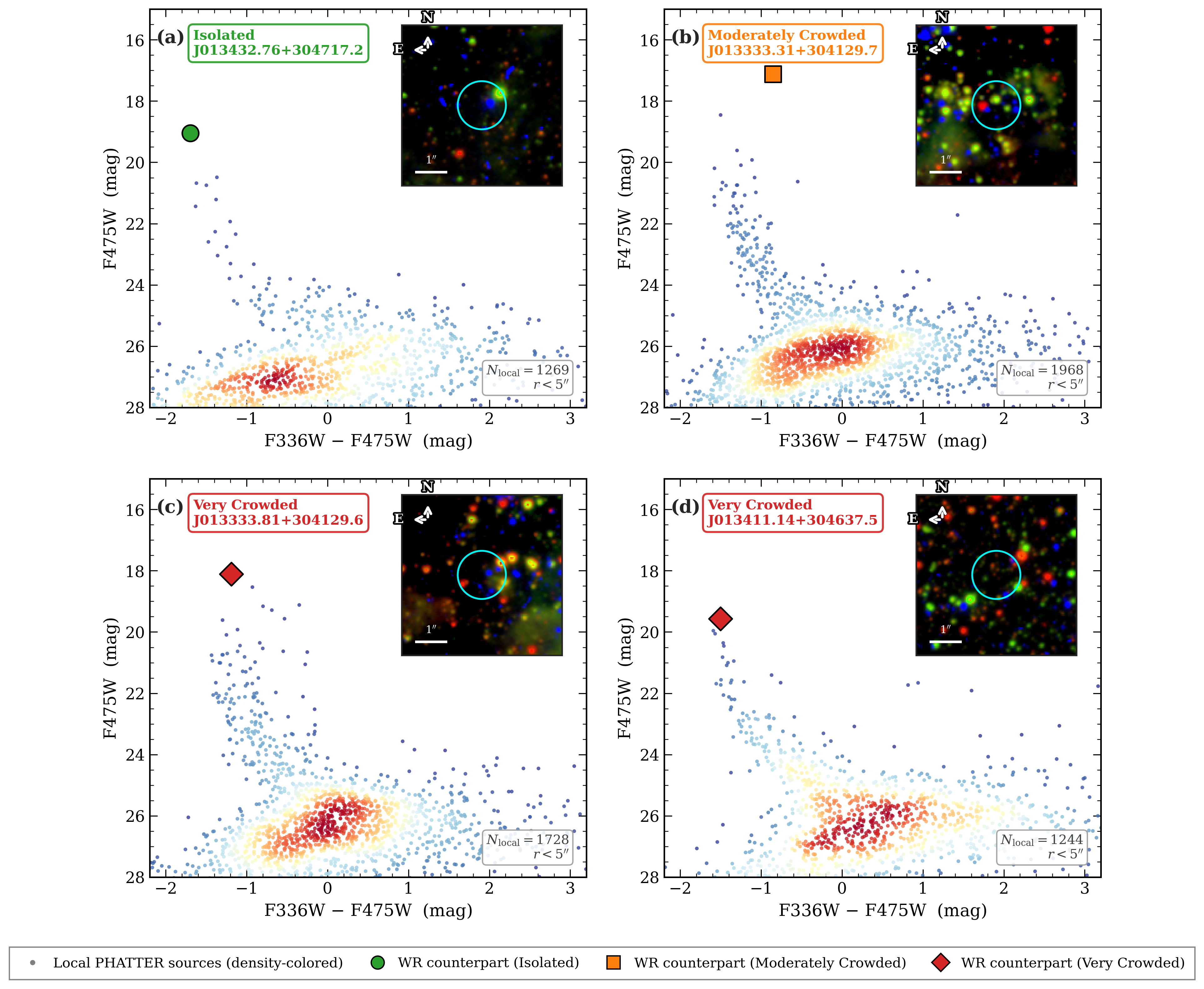}
\caption{Local CMDs (F475W vs.\ F336W$-$F475W) for the four representative
WR stars shown in Figure~\ref{fig:stamps}. Each panel displays all PHATTER
sources within $5\arcsec$ ($\sim$20~pc) of the WR position. Point color
indicates local point-source density from a Gaussian kernel-density estimate,
mapped from low (blue) through intermediate (yellow) to high (red). The
identified WR counterpart (brightest F275W source within $1\arcsec$) is
marked with the corresponding category symbol: green circle (isolated),
orange square (moderately crowded), red diamond (very crowded), as defined in
\S\,\ref{sec:crowding_metric}. \textit{Insets:} $5\arcsec \times 5\arcsec$
HST/PHATTER color composites (blue: F336W; green: F475W; red: F814W) rendered
with a Lupton stretch that preserves relative surface brightness, giving a
direct visual sense of the flux contributed by neighboring sources. The cyan
circle marks the $0\farcs75$-radius Hectospec fiber aperture. Panels: \textbf{(a)}
isolated, \textbf{(b)} moderately crowded, \textbf{(c,\,d)} very crowded.}
\label{fig:local_cmds}
\end{figure*}

\figsetstart
\figsetnum{6}
\figsettitle{HST/PHATTER multiband postage stamps and local source maps for 123 M33 Wolf-Rayet stars with crowding classifications.}

\figsetend

\begin{figure*}[htb!]
\centering
\includegraphics[width=\textwidth, height=0.90\textheight, keepaspectratio]{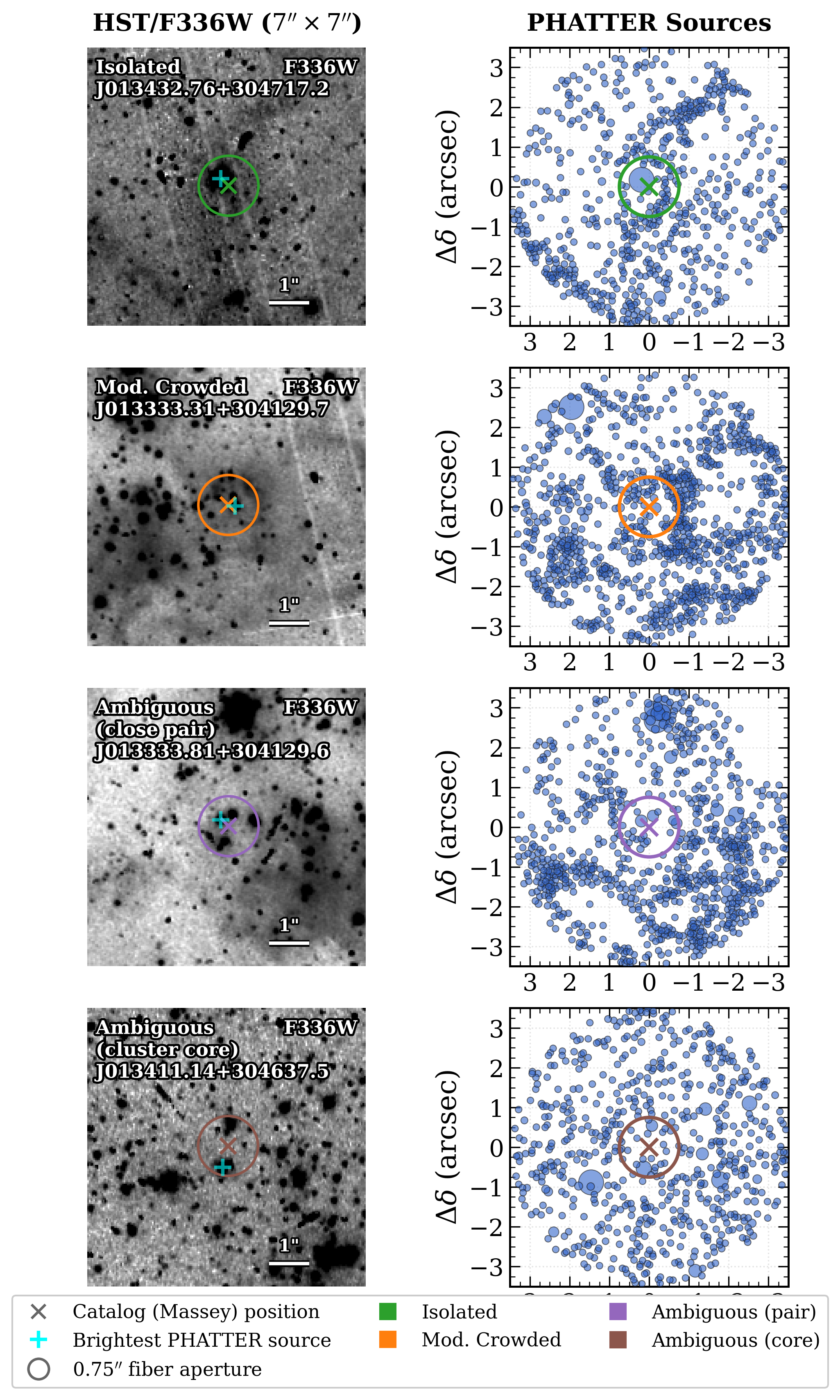}
\caption{Representative \textit{HST}/F336W postage stamps 
($7\arcsec \times 7\arcsec$) and PHATTER source maps for 
four WR stars illustrating the three crowding categories 
defined in \S\,\ref{sec:crowding_metric}. We display F336W imaging because it provides deeper, higher signal-to-noise photometry than F275W while preserving the ultraviolet contrast between hot WR stars and their cooler neighbors. Each row shows one 
WR star: the left panel displays the F336W image with the 
$0\farcs75$-radius synthetic fiber aperture (colored circle), 
the published ground-based catalog position ($\times$), and 
the brightest PHATTER counterpart ($+$); the right panel 
shows all PHATTER sources within $3\farcs5$, with marker 
area proportional to F336W flux. \textit{Row~1 (green):} An 
isolated WR star where a single bright source dominates the 
fiber flux. \textit{Row~2 (orange):} A moderately crowded 
case where a second source contributes measurable flux 
within the fiber. \textit{Row~3 (purple):}  A very crowded ambiguous close pair where two comparably bright blue sources lie within the fiber; the WR spectrum cannot be assigned to either star from photometry alone. \textit{Row~4 (red):} A very crowded case embedded in a dense compact association 
(ambiguous cluster core; \S\,\ref{sec:amb_cases}).}
\label{fig:stamps}
\end{figure*}

\subsection{Color-Magnitude Diagram Analysis} \label{sec:cmd_analysis} 
We constructed global and local CMDs to place the WR stars
in evolutionary context and to provide an independent check
on the crowding classification.

Figure~\ref{fig:global_cmds} shows the WR sample overplotted
on the full PHATTER Hess diagram, color-coded by crowding
category. The isolated sample (green) occupies the expected WR locus---luminous
and extremely blue (F336W$-$F475W $\lesssim -0.3$), consistent with
$T_{\rm eff} \gtrsim 30{,}000$~K \citep{Crowther07}. The moderately and very crowded samples largely share this locus, confirming that the F275W-selected
counterparts are genuine hot sources rather than photometric artifacts.

\rev{A small subset of the very crowded sample ($\sim$3 stars) appears at fainter F336W magnitudes and/or redder F336W$-$F475W colors. The redder colors are consistent with additional flux from cooler neighboring stars, which contribute relatively more at F475W than at F275W, while the fainter F336W magnitudes may reflect intrinsically fainter counterparts rather than crowding. Overall, the three crowding categories overlap substantially in CMD space. CMD position alone therefore provides limited leverage for identifying contaminated systems, compared with the spatial crowding metric in Figure~\ref{fig:crowding}.}

For each target we also generated a local CMD comprising all
PHATTER sources within $5\arcsec$ ($\sim$20~pc) in five
filter combinations spanning the NUV through NIR
(Figure~\ref{fig:local_cmds}).

\begin{table*}[ht!]
\centering
\caption{Serendipitously Observed Wolf-Rayet Stars}
\label{tab:keck_wrs}
\begin{tabular}{l c c c c l}
\toprule
\toprule
Object Name & RA & Dec & V & $v_{\text{helio}}$ & Spectral Type \\
(J2000) & (deg) & (deg) & (mag) & (km s$^{-1}$) & \\
\midrule
J013327.26+303909.1 & 23.3634 & +30.6525 & 17.954 & $-124.0$ & Ofpe/WN9 \\
J013332.97+304136.1 & 23.3873 & +30.6933 & 18.145 & ... & WNL \\
J013334.04+304117.2 & 23.3917 & +30.6881 & 19.779 & $-37.2$ & WN \\
J013353.60+303851.6 & 23.4732 & +30.6477 & 18.091 & $-168.4$ & Ofpe/WN9 \\
J013418.74+303411.8 & 23.5780 & +30.5699 & 19.578 & $-102.7$ & Ofpe/WN \\
\bottomrule
\end{tabular}
\par\medskip
\small{\textit{Note.} Coordinates are J2000. Heliocentric radial velocities ($v_{\text{helio}}$) are derived from Keck/DEIMOS spectra. V-band magnitudes are from the catalog of Massey et al. (2006), updated in Neugent \& Massey (2011, 2014).}
\end{table*}

\subsection{Ambiguous Cases}\label{sec:amb_cases}
Our crowding analysis identified $\sim$10 targets for which the WR counterpart cannot be uniquely determined from broad-band imaging. These cases arise when two or more comparably bright ($\Delta m_{12} \lesssim 0.5$~mag) blue sources lie within the synthetic fiber aperture, each individually consistent with WR-type luminosities. In such configurations the fractional flux contributions are nearly degenerate and the ground-based centroid lies between competing candidates rather than on any individual source. 

These systems fall into two categories: 
\begin{enumerate}

\item \textbf{Close pairs:} Two luminous blue stars separated by $0\farcs2$--$0\farcs5$ ($\sim$1--2~pc), each consistent with WR absolute magnitudes. In some cases both may be genuine WR stars---compact WR+WR or WR+O systems are known in the Milky Way and LMC \citep{Crowther07, Shenar21}. 

\item \textbf{Dense cluster cores:} Targets embedded in the unresolved cores of compact young clusters or giant \ion{H}{2} regions (e.g., NGC~604), where three or more comparable-brightness sources contribute to the fiber flux and even HST imaging cannot isolate the spectroscopic target.
\end{enumerate} 

\noindent Of the $\sim$10 ambiguous targets, roughly 4--6 fall into the close-pair category and the remainder are embedded in dense cluster cores. These targets are flagged in Table~\ref{tab:wr_catalog} and illustrated in the postage-stamp figure (Figure~\ref{fig:stamps}). They are high-priority targets for AO-fed IFU spectroscopy; Keck/OSIRIS (LGS/NGS-AO) can spatially deblend these systems at $\sim0\farcs05$--$0\farcs1$ resolution, substantially finer than the seeing-limited $\sim0\farcs3$--$0\farcs7$ achievable with non-AO IFUs.

\begin{deluxetable*}{lcccccc}
\tablecaption{Spectroscopic Properties of the Five TREX 
Wolf-Rayet Stars\label{tab:line_props}}
\tablewidth{0pt}
\tablehead{
\colhead{Object} &
\colhead{Type} &
\colhead{Line} &
\colhead{EW} &
\colhead{FWHM} &
\colhead{$v_{\rm blue}$\tablenotemark{a}} &
\colhead{$v_{\rm helio}$} \\
\colhead{} &
\colhead{} &
\colhead{} &
\colhead{(\AA)} &
\colhead{(km\,s$^{-1}$)} &
\colhead{(km\,s$^{-1}$)} &
\colhead{(km\,s$^{-1}$)}
}
\startdata
J013334.04+304117.2 & WN
  & \ion{He}{2}\,$\lambda$5411 & $+8.3$ & $1025 \pm 38$ & \nodata & $-37$ \\
J013332.97+304136.1 & WNL
  & \ion{He}{2}\,$\lambda$5411 & $+9.2$ & $682 \pm 14$ & $1504$\tablenotemark{b} & \nodata \\
J013418.74+303411.8 & Ofpe/WN
  & \ion{He}{2}\,$\lambda$4686 & $+35.6$ & $562 \pm 11$ & $755$\tablenotemark{b} & $-103$ \\
J013327.26+303909.1 & Ofpe/WN9
  & \ion{He}{2}\,$\lambda$4686 & $-3.1$\tablenotemark{c} & $389 \pm 67$ & $702$\tablenotemark{b} & $-124$ \\
  &
  & \ion{He}{1}\,$\lambda$5876 & $+6.9$ & $432 \pm 24$ & \nodata & \\
J013353.60+303851.6 & Ofpe/WN9
  & \ion{He}{2}\,$\lambda$4686 & $-3.6$\tablenotemark{c} & \nodata & \nodata & $-168$ \\
  &
  & \ion{He}{1}\,$\lambda$5876 & $+2.8$ & $498 \pm 27$ & \nodata & \\
\enddata
\tablenotetext{a}{\rev{Velocity measured from the blue edge
of the \ion{He}{1} P~Cygni absorption trough, where
detected. $v_{\rm blue}$ from
\ion{He}{1}\,$\lambda$5876 should be treated with caution
for WN subtypes, where broad emission wings can mimic
P~Cygni absorption features.}}
\tablenotetext{b}{P~Cygni absorption detected.}
\tablenotetext{c}{Negative EW indicates net absorption; the 
\ion{He}{2}\,$\lambda$4686 line is in absorption for the 
Ofpe/WN9 subtype, consistent with their transitional 
spectral classification near the O/WN boundary.}
\tablecomments{EWs and FWHM are from Gaussian fits to 
continuum-normalized Keck/DEIMOS spectra. FWHM is converted 
to velocity via $v_{\rm FWHM} = c\,\Delta\lambda/\lambda_0$. 
The WN and WNL objects lack clean coverage at 
\ion{He}{2}\,$\lambda$4686 in the extracted spectra; 
\ion{He}{2}\,$\lambda$5411 is used instead. Heliocentric 
velocities are from the cross-correlation pipeline 
(Table~\ref{tab:keck_wrs}).}
\end{deluxetable*}

\section{Serendipitous Keck/DEIMOS Spectroscopy of Five Wolf-Rayet Stars}\label{sec:Keck_WRs}
 
During our systematic visual inspection of $\approx$7000 TREX spectra (\S\,\ref{sec:keck_spec}), we identified five previously cataloged WR stars that had been serendipitously placed on DEIMOS slitmasks. These objects---all classified as WN-type stars in the \citet{Neugent11, Neugent14} catalog---were not deliberately targeted as WR stars but were selected by the TREX color-magnitude criteria as luminous blue point sources in the PHATTER photometry. Their WN nature was immediately apparent from the strong, broad emission features in the extracted one-dimensional spectra, characteristic of dense, fast stellar winds \citep{Crowther07, Shenar24}. The five targets and their basic properties are listed in Table~\ref{tab:keck_wrs}.
 
Because the TREX target selection preferentially chose relatively isolated sources based on \textit{HST} imaging, and because the Keck observations were obtained in sub-arcsecond seeing ($\lesssim 0\farcs8$), these spectra are substantially less contaminated by neighboring OB stars than the original Hectospec fiber spectra from which the WN classifications were derived. This makes them valuable modern benchmarks for the M33 WN population: while the broad WN classifications agree with those of \citet{Neugent11, Neugent14}, the reduced continuum contamination enables more reliable measurements of individual spectral diagnostics, including emission-line equivalent widths, line widths, and blue-edge velocities from P~Cygni profiles
(\S\,\ref{sec:em_line_widths}).

\subsection{Spectral Classification Confirmation}\label{sec:spec_class}
 
\begin{figure}[htbp!]
    \centering
    \includegraphics[width=1\linewidth]{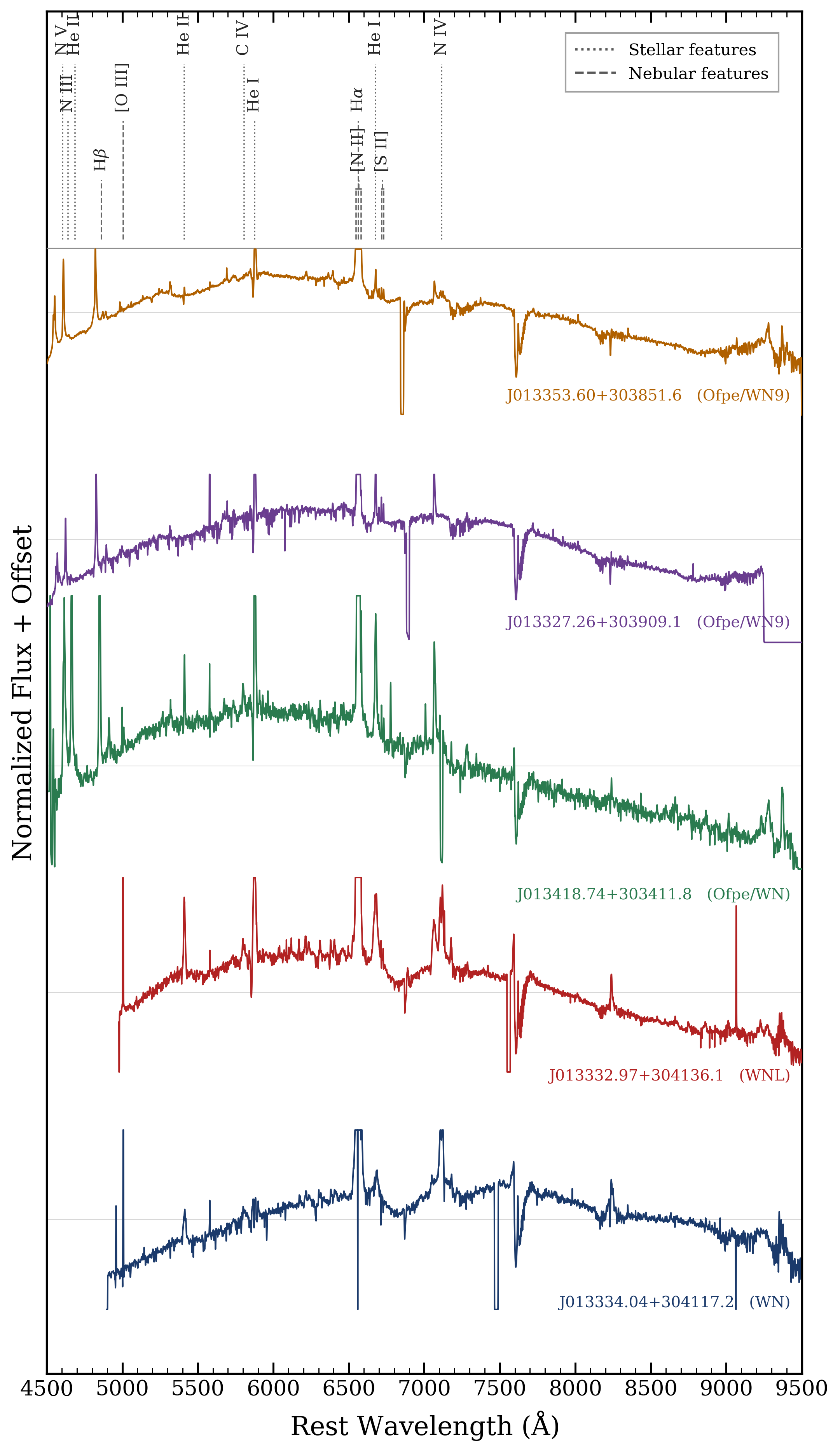}
    \caption{Rest-frame Keck/DEIMOS spectra of the five WN-type 
    Wolf-Rayet stars serendipitously observed during the TREX survey 
    of M33, obtained with the 600ZD grating 
    ($R \approx 2000$; $\lambda\lambda$\,4500--9500\,\AA). Spectra 
    are continuum-normalized, offset vertically for clarity, and 
    ordered by spectral subtype from earliest, hottest winds (bottom) 
    to latest, coolest winds (top). Stellar wind features are marked 
    with dotted lines and nebular emission features with dashed lines. 
    The \ion{He}{2}\,$\lambda5411$ broad emission lines---one of the primary early WN diagnostics---are prominent in the WN and WNL objects but weaken significantly or vanish in the late-type 
    Ofpe/WN9 stars. Conversely, the two Ofpe/WN9 stars (top) exhibit 
    narrower emission profiles and significant \ion{He}{1} features 
    (notably the P~Cygni profile at \ion{He}{1}\,$\lambda5876$) 
    relative to the earlier WN subtype (bottom), consistent with 
    their cooler, slower, and denser winds. The progression from 
    broad, high-ionization features (WN) to narrow, low-ionization 
    profiles (Ofpe/WN9) confirms the published WN classifications 
    and, through reduced fiber contamination, enables quantitative 
    wind diagnostics inaccessible from the original Hectospec data 
    (\S\,\ref{sec:em_line_widths}, 
    \S\,\ref{sec:Halpha_decontam}).}
    \label{fig:TREX_WR_Spectra}
\end{figure}

Figure~\ref{fig:TREX_WR_Spectra} presents the continuum-normalized Keck/ DEIMOS spectra for the five WN-type stars in our sample, arranged in a spectral ladder spanning $\lambda\lambda$4500--9500~\AA. The sample illustrates the clear spectral progression of the WN sequence. The earlier, hotter WN and WNL subtypes (bottom traces) are dominated by strong, broad \ion{He}{2}\,$\lambda$5411 emission, indicative of fast, highly ionized stellar winds. Conversely, the two objects at the cool end of the sequence---J013327.26+303909.1 and J013353.60+303851.6 (top traces)---display the relatively narrow emission lines and \ion{He}{1}-dominated spectra characteristic of the Ofpe/WN9 transitional class. In these cooler winds, the \ion{He}{2} features weaken significantly or vanish entirely, while \ion{He}{1} lines (e.g., $\lambda$5876) develop distinct P-Cygni absorption troughs. These morphologies are consistent with their published catalog subtypes.
We note that none of the five spectra show detectable \ion{C}{4}~$\lambda5808$ emission, confirming the absence of WC characteristics and ruling out WN/WC transitional status.

Three of the five WR stars in our sample were 
re-observed with Keck/DEIMOS during follow-up 
programs in 2022 and 2023, providing a temporal 
baseline of 4--5~years relative to the original 
2018 TREX observations. \rev{The follow-up 
observations used the 950G grating 
($\lambda\lambda$\,4800--10\,000\,\AA), which does 
not cover the primary blue classification 
diagnostics (\ion{He}{2}\,$\lambda$4686, 
\ion{N}{3}\,$\lambda$4640), so the multi-epoch 
comparison is limited to 
\ion{He}{1}\,$\lambda\lambda$5876,\,7065 and 
\ion{He}{2}\,$\lambda$5411. The equivalent widths 
of these features for the two Ofpe/WN9 stars are 
consistent within $\lesssim 0.5$\,\AA\ between 
the 2018B and 2022B/2023B epochs. We assessed 
radial-velocity stability by cross-correlating the 
\ion{He}{1}\,$\lambda7065$ profile between 
same-grating (950G) epochs to avoid wavelength 
zero-point offsets between grating configurations. 
For J013353.60$+$303851.6 (Ofpe/WN9), observed in 
both 2022B and 2023B, no measurable velocity shift 
is detected over the one-year baseline, consistent 
with the instrumental precision of $\sim$30~km\,s$^{-1}$ at $R \approx 2000$. Sparse multi-epoch sampling does not rule out long-period or low-amplitude binarity.}




\subsection{Nebular H$\alpha$ Decontamination}\label{sec:Halpha_decontam}
\begin{figure*}[htb!]
\centering
\includegraphics[width=\textwidth]{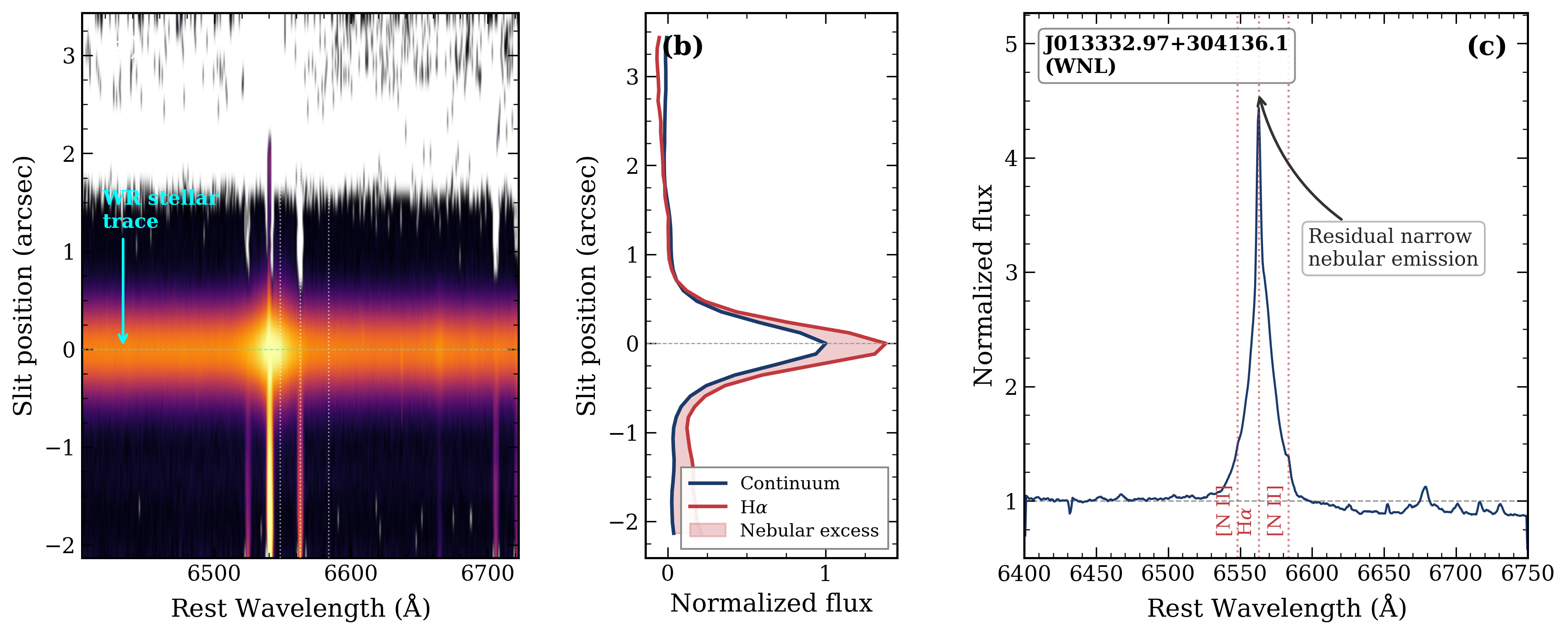}

\caption{Nebular H$\alpha$ contamination in the DEIMOS slit spectrum of J013332.97$+$304136.1 (WNL). 
\textbf{(a)}~Two-dimensional slit image centered on H$\alpha$; the color scale maps flux from low (dark) to high (bright). The compact horizontal trace is the stellar continuum; the vertical features extending along the slit are the [\ion{N}{2}]\,$\lambda\lambda$6548,\,6583 and H$\alpha$ nebular emission lines (white dotted lines mark their rest wavelengths). The cyan arrow indicates the WR stellar trace. 
\textbf{(b)}~Spatial profiles at H$\alpha$ ($\pm$12\,\AA; red) and in adjacent line-free continuum bands (6350--6450\,\AA\ and 6650--6750\,\AA; blue), both normalized to the continuum peak. The shaded region marks the excess of the H$\alpha$ profile over the stellar PSF. 
\textbf{(c)}~One-dimensional extracted spectrum. The dashed horizontal line marks the continuum level; red dotted vertical lines and labels identify the [\ion{N}{2}] and H$\alpha$ rest wavelengths. The narrow spike at the H$\alpha$ rest wavelength is residual nebular emission that spatially overlaps the stellar PSF and survives local background subtraction. This target shows the largest nebular H$\alpha$ fraction among the four slit-observed WR stars (\S\,\ref{sec:Halpha_decontam}).}
\label{fig:Halpha_decontam}
\end{figure*}

Each DEIMOS slit extends several arcseconds along the spatial axis,
enabling separation of the spatially extended narrow nebular
H$\alpha$ emission ($\mathrm{FWHM}\sim30$--50~km\,s$^{-1}$) from
the spatially unresolved broad stellar-wind component
($\mathrm{FWHM}\gtrsim1000$~km\,s$^{-1}$). The standard
\texttt{spec2d} sky-subtraction procedure \citep{cooper2012spec2d}
models the extended emission from off-source rows and removes it
from the extracted one-dimensional spectrum.
Figure~\ref{fig:Halpha_decontam} illustrates this for
J013332.97$+$304136.1 (WNL), the most severely contaminated target,
where the narrow nebular component contributes $\approx$80\% of the
total H$\alpha$ flux at the stellar position. Across the four
slit-observed targets with adequate H$\alpha$ coverage, nebular
fractions range from a few percent to $\sim$80\%, reflecting
differences in local \ion{H}{2}-region surface brightness.
\rev{At sufficient spectral resolution ($R \gtrsim 1000$), the
narrow nebular component can generally be distinguished spectrally;
spatially resolved spectroscopy provides an additional, independent
means of separation.}

 
\subsection{Emission Line Properties}
\label{sec:em_line_widths}

We measured equivalent widths, line widths, and
wind velocities from the continuum-normalized
Keck/DEIMOS spectra; results are summarized in
Table~\ref{tab:line_props}.

We fitted single Gaussian profiles to the strongest
\ion{He}{2} feature available for each star:
\ion{He}{2}\,$\lambda5411$ for the WN and WNL
objects, and \ion{He}{2}\,$\lambda4686$ for the
Ofpe/WN and Ofpe/WN9 subtypes. The measured
$v_{\rm FWHM}$ values range from
329--1025~km\,s$^{-1}$, with the Ofpe/WN9 objects
exhibiting systematically narrower profiles
($v_{\rm FWHM} \approx 390$--$560$~km\,s$^{-1}$)
than the WNL (682~km\,s$^{-1}$) and WN
(1025~km\,s$^{-1}$) subtypes, consistent with the
known progression from slow Ofpe/WN9 outflows to
fast WNE winds \citep{Crowther07}. The two Ofpe/WN9
stars show \ion{He}{2}\,$\lambda4686$ in net
absorption ($\mathrm{EW} \approx -3$\,\AA),
confirming their position at the O/WN boundary
\citep{Mathys88}.

For lines exhibiting P~Cygni profiles, we measured
\rev{the blue-edge velocity $v_{\rm blue}$} from
the blue edge of the absorption trough, defined as
the wavelength at which the profile rises back to
within $1\sigma$ of the local continuum. P~Cygni
profiles were identified in
\ion{He}{1}\,$\lambda5876$ for three of the five
targets, yielding \rev{$v_{\rm blue} =
702$--$1504$~km\,s$^{-1}$}
(Table~\ref{tab:line_props}).
\rev{Because \ion{He}{1}\,$\lambda5876$ is a
subordinate line and the absorption troughs are
unsaturated, these blue-edge velocities do not
necessarily correspond to the true $v_\infty$;
we therefore report them as \ion{He}{1} blue-edge
velocities throughout. For hotter WN subtypes,
where \ion{He}{1} is a minority ion, broad
\ion{He}{2} emission wings can mimic P~Cygni
morphology, and the measured $v_{\rm blue}$ values
should be treated as upper limits.}
 
\begin{figure*}[htb!]
\centering
\includegraphics[width=\textwidth]{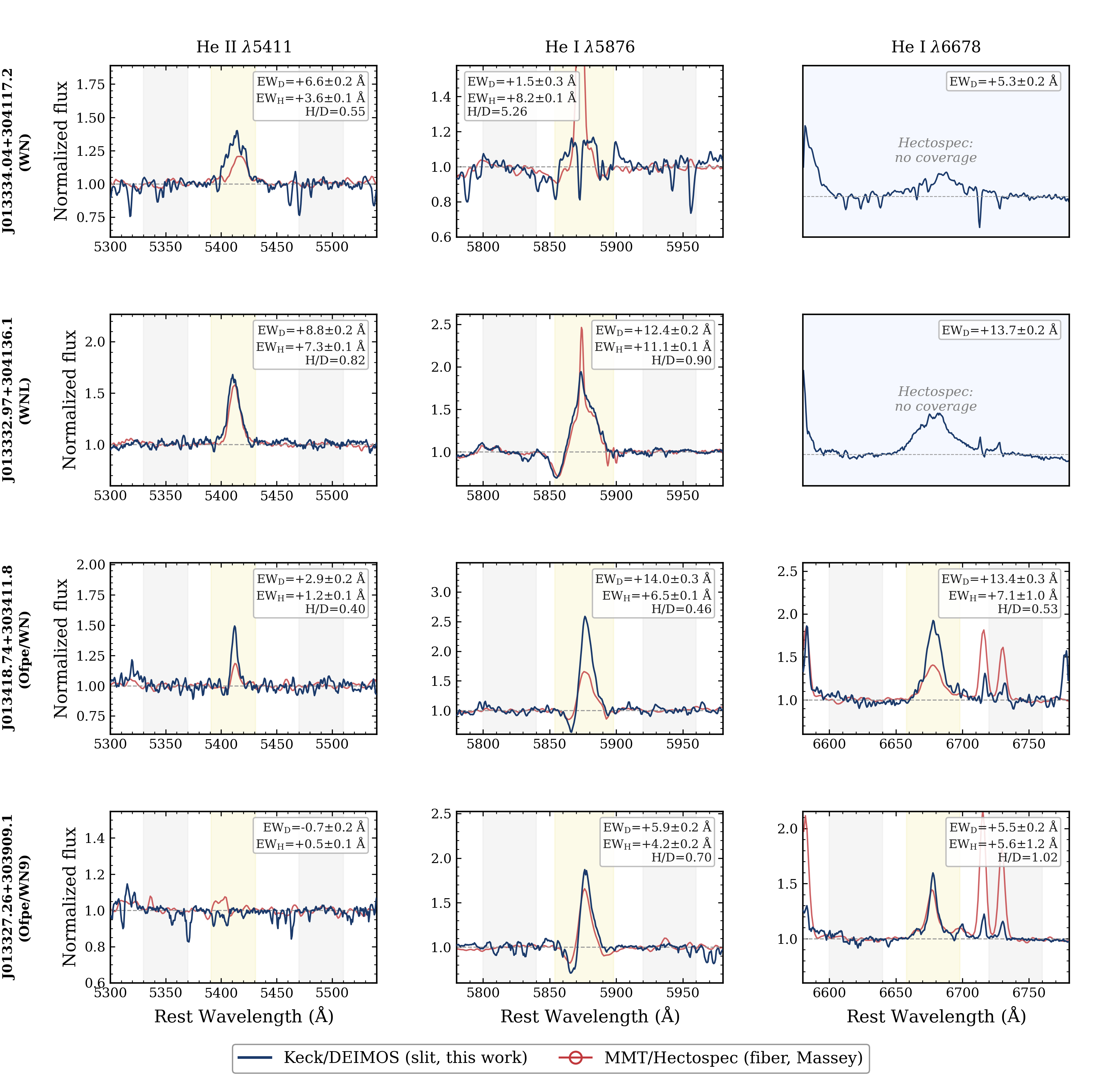}
\caption{Continuum-normalized Keck/DEIMOS (blue; slit, this work) and 
MMT/Hectospec (red; $1\farcs5$ fiber, Massey et al.) spectra for the 
four WR stars, zoomed on three wind 
diagnostic regions. Each row corresponds to one target (labeled at 
left with spectral subtype); columns show 
\ion{He}{2}\,$\lambda5411$, \ion{He}{1}\,$\lambda5876$, and 
\ion{He}{1}\,$\lambda6678$. Both spectra are continuum-normalized 
using linear fits to the grey-shaded sideband windows; the 
yellow-shaded region marks the EW integration window. Measured 
equivalent widths ($\mathrm{EW}_{\rm D}$, $\mathrm{EW}_{\rm H}$) 
and the Hectospec-to-DEIMOS ratio H/D are annotated in each panel. 
Panels marked ``Hectospec: no coverage'' indicate that the Hectospec 
wavelength solution does not extend to \ion{He}{1}\,$\lambda6678$ for those observations. \rev{From top to bottom: 
J013334.04$+$304117.2 (WN; Very Crowded, 
$F_{\rm WR}^{475} = 0.80$), 
J013332.97$+$304136.1 (WNL; Moderately Crowded, 
$F_{\rm WR}^{475} = 0.75$), 
J013418.74$+$303411.8 (Ofpe/WN; Isolated, 
$F_{\rm WR}^{475} = 0.97$), and 
J013327.26$+$303909.1 (Ofpe/WN9; Moderately Crowded, 
$F_{\rm WR}^{475} = 0.83$).}}
\label{fig:ew_comparison}
\end{figure*}


\section{Discussion}
\label{sec:discussion}

Our HST/PHATTER analysis quantifies, for the first time, the photometric contamination affecting individual M33 WR stars and provides Keck/DEIMOS slit spectra free from the fiber-blending and nebular contamination that has characterized previous ground-based work. In this section we discuss the implications of these results for spectral classification (\S\,\ref{sec:disc_class}), wind diagnostics 
(\S\,\ref{sec:disc_winds}), and the broader use of M33 as a calibration anchor for population synthesis (\S\,\ref{sec:disc_popsyn}), and we outline the catalog's 
practical utility (\S\,\ref{sec:disc_utility}).

\subsection{Impact of Crowding on Spectral Classification}
\label{sec:disc_class}

We find that 33 of 123 WR stars in the PHATTER footprint ($\sim$27\%) are classified as moderately or very crowded by our two-parameter metric
(\S\,\ref{sec:crowding_metric}), with $F_{\rm WR}$ values typically below $\sim$0.85 and $\sim$0.50 for the two subsamples, respectively. The sensitivity of the WN classification system to this dilution depends on which diagnostic is used.

\textit{Line-ratio classifications are largely preserved.}
The WN subtype sequence within the WNE and WNL classes is defined by ratios such as \ion{N}{3}\,$\lambda$4640/\ion{He}{2}\,$\lambda$4686 and
\ion{He}{1}\,$\lambda$5876/\ion{He}{2}\,$\lambda$5411 \citep{Smith96, Crowther07}. Under continuum dilution, both numerator and denominator scale equally with $F_{\rm WR}$, so line ratios are invariant.

\textit{Absolute EW thresholds, however, are directly affected.}
The boundary between WNL and the Ofpe/WN9 transitional class sits at
EW(\ion{He}{2}\,$\lambda$4686) $\approx 10$--$15$\,\AA\ \citep{Crowther07, Mathys88}. A WNL star with an intrinsic EW of 30\,\AA, typical of the class \citep{Hamann19}, would appear at 12\,\AA\ when observed through a fiber with $F_{\rm WR}=0.4$, marginally crossing into the Ofpe/WN9 regime. Our
moderately and very crowded samples are therefore the most susceptible to systematic WNL\,$\rightarrow$\,Ofpe/WN9 misclassification. A related effect was reported by \citet{Bibby10} for WR stars in starburst galaxies observed through
ground-based seeing.

Our DEIMOS--Hectospec EW comparison (Figure~\ref{fig:ew_comparison}) provides a direct empirical test. For the \ion{He}{2} and \ion{He}{1}\,$\lambda6678$ diagnostics the Hectospec EW is systematically smaller than the DEIMOS value, in the sense predicted by continuum dilution. Excluding the two cases discussed below, the measured ratios span ${\rm H/D} = 0.55$--$0.90$, comparable in magnitude to the $F_{\rm WR}^{475}$ values of the same targets. The agreement is
not exact. J013332.97$+$304136.1 (WNL) shows ${\rm H/D} = 0.90 \pm 0.03$ at \ion{He}{2}\,$\lambda5411$ against $F_{\rm WR}^{475} = 0.75$, while J013334.04$+$304117.2 (WN) shows ${\rm H/D} = 0.55 \pm 0.03$ against $F_{\rm WR}^{475} = 0.80$. The two deviations are of opposite sign and comparable size. With only four stars spectra available for direct comparison, we can confirm the direction and approximate magnitude of the dilution but cannot test the quantitative
prediction ${\rm H/D} = F_{\rm WR}$. Individual $F_{\rm WR}$ values in
Table~\ref{tab:wr_catalog} should therefore be treated as estimates of the dilution, not as bounds on it. A larger slit-spectroscopic sample is needed to determine whether the residual scatter reflects measurement uncertainty, the adopted seeing model, or low-level intrinsic variability.

J013418.74$+$303411.8 (Ofpe/WN) has the highest $F_{\rm WR}^{475} = 0.97$ in the comparison sample, yet shows ${\rm H/D} = 0.40$--$0.53$ across the three measured lines (Figure~\ref{fig:ew_comparison}). Continuum dilution cannot produce this. The Hectospec and DEIMOS epochs are separated by roughly a
decade, and Ofpe/WN stars are known to vary on this timescale. The M33
catalog includes at least one object with an Ofpe/WN9 spectrum that is a
confirmed LBV (Var\,2). Intrinsic emission-line variability is the most likely explanation, though the low Hectospec S/N for this faint target ($V = 19.6$) may also contribute. Second, the \ion{He}{1}\,$\lambda5876$ measurement for J013334.04$+$304117.2 gives ${\rm H/D} = 5.26$, with the Hectospec EW exceeding the DEIMOS value. The DEIMOS EW at this line is small ($1.5 \pm 0.3$\,\AA), so the ratio is poorly constrained, and the Hectospec profile shows a narrow component absent from the spatially resolved DEIMOS slit extraction. Because this star has the lowest nebular fraction in our sample (\S,\ref{sec:Halpha_decontam}), we attribute the excess to a blended narrow feature in the fiber rather than to nebular recombination or the WR wind.

The dilution bias has a further consequence for the inferred surface hydrogen mass fraction $X_H$, which discriminates between hydrogen-burning WNh stars ($M_{\rm init}\gtrsim80$--$100\,M_\odot$) and classical, hydrogen-depleted cWN stars \citep{Grafener08, Shenar21, Shenar24}. Continuum dilution weakens the broad \ion{He}{2}\,$\lambda$4686 emission while preserving, or enhancing,
hydrogen features contributed by OB neighbors. This artificially inflates $X_H$ and can mask the true evolutionary state of the star.

Our crowding classification enables a first-order correction. For the isolated sample ($N=90$), the classifications of \citet{Neugent11, Neugent14} can be considered reliable. For the moderate sample ($N=13$), published subtypes should be treated as lower limits on the true emission-line strength, with reclassification to earlier subtypes plausible after dilution correction. For the very crowded sample ($N=20$), spectral types are unreliable and spatially resolved follow-up is required for definitive classification.

\subsection{Implications for Wind Diagnostics and 
Mass-Loss Rates}
\label{sec:disc_winds}
For emission-line-based $\dot{M}$ estimates, crowding
introduces a systematic underestimate. Equivalent widths
are diluted by $F_{\rm WR}$; combined with WR-only
continuum fluxes from HST photometry, this yields line
luminosities underestimated by the same factor. Because
recombination-line luminosities scale as $\dot{M}^2$ in
the optically thin regime \citep{Crowther07, Hamann06},
an $F_{\rm WR}$ underestimate of $L_{\rm line}$
translates to an $F_{\rm WR}^{1/2}$ underestimate of
$\dot{M}$ \rev{when the bolometric luminosity and other
stellar parameters are held fixed or independently
constrained (e.g., from the deblended HST photometry
provided here). If the bolometric luminosity is itself
contaminated by neighbor flux, the corresponding bias in
$R_*$ can partially compensate the EW effect through the
transformed-radius scaling, and the net $\dot{M}$ bias
may therefore be smaller.} For our very crowded sample
($F_{\rm WR} \sim 0.3$), the $F_{\rm WR}^{1/2}$ estimate
corresponds to a $\sim$45\% bias \rev{in the
fixed-luminosity limit}. The F475W- and F814W-band $F_{\rm WR}$ 
values tabulated in the catalog Table~\ref{tab:wr_catalog} provide 
the appropriate correction factors for optical diagnostic 
lines (\S\,\ref{sec:synth_fiber}).
Critically, because Hectospec is not flux-calibrated 
\citep[][, \S\,\ref{sec:Keck_WRs}]{Neugent11}, the 
absolute line fluxes that would in principle survive 
continuum dilution were never available for the 
foundational M33 classifications.

\rev{Nebular H$\alpha$ contamination introduces a 
second bias in fiber-based $\dot{M}$ estimates 
(\S\,\ref{sec:Halpha_decontam}). In practice, WR 
atmosphere models generally account for nebular 
contamination by excluding narrow nebular features 
from the fit, although this becomes more difficult 
when nebular and wind components overlap.} The 
transformed radius 
$R_t \propto R_*(v_\infty/\dot{M})^{2/3}$ 
\citep{Schmutz89, Hamann06} depends on both the 
stellar luminosity (and hence $R_*$) and $\dot{M}$. 
The six-band HST photometry provided here enables 
WR-only luminosity estimates and, together with the 
$F_{\rm WR}$ correction factors, reduces 
continuum-dilution systematics in future atmospheric 
fitting.

\rev{Our Keck/DEIMOS spectra provide empirical
\ion{He}{1} blue-edge velocity measurements
($v_{\rm blue} \approx 700$--$1500$\,km\,s$^{-1}$;
Table~\ref{tab:line_props}). These measurements
are less affected by continuum dilution than
EW-based diagnostics, although they should not be
interpreted as terminal wind velocities
(\S\,\ref{sec:em_line_widths}). They provide useful
constraints for future detailed atmosphere modeling
with \textsc{PoWR} or \textsc{cmfgen}.}

\subsection{Catalog Utility and Future Prospects}
\label{sec:disc_utility}

The clean sample of 90 WR stars with robust six-band 
HST photometry constitutes a high-fidelity training set 
for machine-learning classifiers identifying rare 
stellar types in large photometric surveys, while our 
flags for crowded and ambiguous sources can filter 
contaminated targets that would otherwise degrade 
automated classifications. Observational follow-up programs can preferentially target our isolated sample for clean wind diagnostics with seeing-limited IFUs or JWST/NIRSpec (where the 
higher spatial resolution mitigates nebular 
contamination relative to fiber spectroscopy), or 
our ambiguous and very crowded subsamples for 
definitive deblending via AO-fed IFU spectroscopy 
with Keck/OSIRIS (LGS/NGS-AO). 

Finally, PHATTER covers only $\sim$50\% of the M33 WR 
population. An extension to larger galactocentric radii 
would bring the remaining 89 outer-disk WR stars within 
reach of this analysis, where M33's metallicity falls 
to $\sim$0.3\,Z$_\odot$ \citep{Bresolin11}---precisely 
the regime where theoretical mass-loss predictions 
diverge most sharply \citep{Vink01}.

\section{Summary}\label{sec:summary}

We have presented a systematic crowding analysis of the M33 
WR population using HST/PHATTER imaging, combined 
with new Keck/DEIMOS slit spectroscopy of five serendipitously 
observed WR stars. Our main results are:

\begin{enumerate}

\item We developed a two-parameter crowding diagnostic 
($d$, $\Delta m_{12}$; \S\,\ref{sec:crowding_metric}) 
and a complementary synthetic-fiber model 
(\S\,\ref{sec:synth_fiber}) that together categorize and quantify 
the fractional WR continuum contribution $F_{\rm WR}$ 
to the $1\farcs5$ Hectospec fiber used for the 
foundational WR census. Applied to the 123 WR stars 
within the PHATTER footprint, the classification 
yields 90 isolated, 13 moderately crowded ($F_{\rm WR}$ typically $\sim$0.4--0.9), and 20 very crowded ($F_{\rm WR}$ typically $\lesssim 0.5$) systems, where the classification is defined by the model-independent $(d, \Delta m_{12})$ metric and $F_{\rm WR}$ is the corresponding fiber-model estimate (\S\,\ref{sec:crowding_metric}). Over one-quarter of the M33 WR population within PHATTER suffers significant continuum dilution 
from unresolved OB neighbors.

\item We provide updated HST coordinates and 
six-band photometry (F275W, F336W, F475W, F814W, 
F110W, F160W) for 123 M33 WR stars, with the WR 
counterpart identified as the brightest F275W source 
within $1\arcsec$ of the catalog position 
(\S\,\ref{sec:crossmatch}). Residual astrometric offsets between the corrected ground-based positions and the HST counterparts scale with crowding class (isolated: median $0\farcs13$; very crowded: median $0\farcs26$, 
maximum $0\farcs88$), providing a geometry-based check consistent with the crowding classifications (Figure~\ref{fig:astrometric_offsets}). The catalog is publicly 
available in machine-readable form 
(Table~\ref{tab:wr_catalog}). The isolated sample 
occupies the expected luminous, blue locus on global 
CMDs, consistent with the crowding 
classification. A subset of very crowded stars is 
displaced toward redder colors, the photometric 
signature of unresolved OB continuum dilution in the 
optical bands. 

\item We identify $\sim$10 ambiguous cases in which 
two or more comparably bright blue sources lie within 
the fiber aperture and the WR counterpart cannot be 
uniquely assigned from broad-band imaging 
(\S\,\ref{sec:amb_cases}). These arise primarily in 
the cores of NGC~604 and NGC~595. Because 
sub-arcsecond spatial resolution is essential for 
deblending these systems, they are high-priority 
targets for laser- or natural-guide-star AO-fed 
IFU spectroscopy; Keck/OSIRIS (LGS/NGS-AO) is 
particularly well-suited given M33's northern 
declination ($\delta \approx +30\degr$), which 
makes it a challenging target for VLT/MUSE.

\item We present new Keck/DEIMOS spectra 
($\lambda\lambda$\,4500--9500\,\AA, $R\approx2000$) 
for five WN-type stars serendipitously observed during 
the TREX survey (\S\,\ref{sec:spec_class}). Direct EW comparison with archival MMT/Hectospec spectra for four of these targets (Figure~\ref{fig:ew_comparison}) confirms the predicted dilution:
${\rm EW_{Hecto}/EW_{DEIMOS}} = 0.55$--$0.90$, with the Hectospec values
uniformly smaller. The sense and approximate magnitude of the effect match the $F_{\rm WR}$ prediction, but four stars are too few to
test the relation ${\rm EW_H/EW_D} = F_{\rm WR}$ quantitatively.

\item We measure \ion{He}{2} line widths 
($v_{\rm FWHM} \approx 330$--$1025$~km\,s$^{-1}$; 
Table~\ref{tab:line_props}) and P~Cygni \rev{blue-edge velocities of \ion{He}{1} absorption($v_{\rm blue} \approx 700$--$1500$~km\,s$^{-1}$)}, 
detected in three of five targets from the 
DEIMOS spectra (\S\,\ref{sec:em_line_widths}). The 
Ofpe/WN9 subtypes show systematically narrower 
profiles than the WN/WNL subtypes, consistent with 
their cooler, denser winds. These measurements provide 
modern wind benchmarks for the M33 WN population 
at its intermediate metallicity 
($Z \approx 0.5$--$1.0\,Z_\odot$), and are robust 
against continuum dilution since they are referenced 
to the local continuum rather than to integrated 
line fluxes.

\item Multi-epoch Keck/DEIMOS spectra spanning 4--5 
years for three of the five targets reveal no 
statistically significant EW or radial-velocity 
variability ($\Delta\mathrm{EW} \lesssim 0.5$~\AA, 
$|\Delta v| \lesssim 30$~km\,s$^{-1}$; 
\S\,\ref{sec:spec_class}). This temporal stability 
argues against short-period binary modulation 
($P \lesssim 10$~yr) and rules out the eruptive 
variability characteristic of LBV impostors, 
reinforcing the classification of these objects as 
classical, stable WR stars.

\item For the four WR stars with clean DEIMOS slit extractions, the two-dimensional spectral image reveals spatially 
extended [\ion{N}{2}] and H$\alpha$ emission from 
the surrounding \ion{H}{2} region, with nebular 
fractions at the stellar position ranging from 
$\lesssim\!5\%$ to $\sim\!80\%$ depending on the 
local nebular environment (\S\,\ref{sec:Halpha_decontam}). 
In MMT/Hectospec fiber spectra, the nebular 
component is blended with the stellar 
wind emission, systematically inflating the apparent 
H$\alpha$ luminosity and biasing H$\alpha$-based 
mass-loss rate determinations. This effect is 
irremovable in fiber spectra but can be spatially 
separated with slit or IFU spectroscopy.

\end{enumerate}

\noindent
The crowding-corrected catalog and the new slit spectra 
together provide a foundational reference for 
re-evaluating M33's role as a calibration anchor for 
population synthesis models, for future spatially 
resolved follow-up of the ambiguous and very crowded 
subsamples, and for the construction of high-fidelity 
photometric training sets for machine-learning WR 
classifiers.

\section*{Data Availability}
\label{sec:data_availability}

The full machine-readable catalog (Table~\ref{tab:wr_catalog}), 
containing \textit{HST}/PHATTER positions, six-band 
photometry (F275W--F160W, VEGAMAG), crowding 
classifications, $F_{\rm WR}$ values, and spectral 
cross-identifications for all 123 WR stars within 
the PHATTER footprint, is available via the journal's 
online supplement and at 
\href{https://github.com/RohitRajAstro/M33-WR-Crowding}{M33-WR-Crowding}, 
which also contains the Keck/DEIMOS one-dimensional 
extracted spectra (FITS format) for the five WR stars 
presented in \S\,\ref{sec:Keck_WRs}. The underlying 
PHATTER source catalogs are publicly available on 
MAST:\dataset[doi:10.17909/t9-ksyp-na40]{https://doi.org/10.17909/t9-ksyp-na40} \citep{Williams21}.\footnote{\url{https://archive.stsci.edu/hlsp/phatter}}

\textit{HST}/F275W, F336W, F475W, and F160W postage 
stamps and local color-magnitude diagrams for all 
123 WR stars are provided as figure sets in the 
online journal (Figures~\ref{fig:stamps} and~\ref{fig:local_cmds}).

\section*{Acknowledgments}
We thank the anonymous referee for a constructive report that improved the clarity of this paper. We thank Philip Massey for helpful discussions, suggestions, and access to Hectospec spectroscopy data from the M33 WR census. We thank Matt Ashby and Andrea Dupree for a careful reading of the manuscript and helpful suggestions that improved the presentation. We thank Olivia Gaunt and Preksha Sethia for their contributions to the early stages of this project. This work is based on observations made with the NASA/ESA \textit{Hubble Space Telescope}, obtained from the Mikulski Archive for Space Telescopes (MAST) at the Space Telescope Science Institute. Support for Program HST GO-14610 was provided by NASA through a grant from the Space Telescope Science Institute under NASA contract NAS5-26555.

The spectroscopic data presented herein were obtained 
at the W.\ M.\ Keck Observatory. The authors wish to recognize and 
acknowledge the very significant cultural role and 
reverence that the summit of Maunakea has always had 
within the indigenous Hawaiian community. We are most 
fortunate to have the opportunity to conduct 
observations from this mountain.

The TREX spectroscopic survey was supported by 
National Science Foundation grants AST-1909066 
(KMG) and AST-1909759 (PG).


\textit{Software:} \texttt{astropy} 
\citep{astropy13, astropy18, astropy22}, 
\texttt{numpy} \citep{numpy20}, 
\texttt{matplotlib} \citep{matplotlib07}, 
\texttt{scipy} \citep{scipy20}, 
\texttt{spec2d}/\texttt{spec1d} \citep{cooper2012spec2d, 
Newman13}.

\textit{Facilities:} HST (WFC3/UVIS, ACS/WFC, WFC3/IR), 
Keck II (DEIMOS), MMT (Hectospec).

\section*{ORCID iDs}
\noindent Rohit Raj \url{https://orcid.org/0009-0000-0170-6302} \\
Puragra Guhathakurta \url{https://orcid.org/0000-0001-8867-4234} \\
Robert E. Williams \url{https://orcid.org/0000-0002-3742-8460} \\
Benjamin F. Williams \url{https://orcid.org/0000-0002-7502-0597} \\
Julianne J. Dalcanton \url{https://orcid.org/0000-0002-1264-2006} \\
Meredith J. Durbin \url{https://orcid.org/0000-0001-7531-9815}

\bibliography{bibtex}
\bibliographystyle{aasjournal}

\end{document}